\documentclass{emulateapj}
\usepackage{epsfig}
\usepackage{color}
\newcommand{\msun}{\,\rm M_\odot}

\newcommand{\kmsmpc}{\,{\rm km\,s^{-1}\,Mpc^{-1}}}
\newcommand{\be}{\begin{equation}}
\newcommand{\ee}{\end{equation}}
\newcommand{\ba}{\begin{eqnarray}}
\newcommand{\ea}{\end{eqnarray}}
\newcommand{\f}{\frac} 
\newcommand{\rvir}{r_{\rm vir}}
\newcommand{\cvir}{c_{\rm vir}}
\newcommand{\mvir}{M_{\rm vir}}
\newcommand{\mhalo}{M_{\rm halo}}
\newcommand{\rtwo}{r_{200}}
\newcommand{\mtwo}{M_{200}}
\newcommand{\msub}{M_{\rm sub}}

\newcommand{\vmax}{V_{\rm max}}
\newcommand{\rvmax}{r_{\rm Vmax}}

\newcommand{\rfm}{r_{\rm M}}
\newcommand{\kpc}{\,{\rm kpc}}
\newcommand{\kms}{\,{\rm km/s}}

\lefthead{Diemand, Kuhlen, \& Madau}
\righthead{Evolution of $\Lambda$CDM halos and their substructure}



\begin{document}
\submitted{}

\title{Formation and evolution of galaxy dark matter halos and their substructure}
\author{J\"urg Diemand\altaffilmark{1,2}, Michael Kuhlen\altaffilmark{3}, \& Piero 
Madau\altaffilmark{1,4}}

\altaffiltext{1}{Department of Astronomy \& Astrophysics, University of
California, Santa Cruz, CA 95064.}
\altaffiltext{2}{Hubble Fellow.}
\altaffiltext{3}{School of Natural Sciences, Institute for Advanced Study,
Einstein Drive, Princeton, NJ 08540.}
\altaffiltext{4}{Max-Planck-Institut f\"ur Astrophysik, Karl-Schwarzschild-Str.
1, 85740 Garching, Germany.}
\email[email: ]{diemand@ucolick.org, mqk@ias.edu, pmadau@ucolick.org.}

\begin{abstract}
We use the ``Via Lactea'' simulation to study the co-evolution of a Milky Way-size $\Lambda$CDM halo
and its subhalo population. While most of the host halo mass is accreted over 
the first 6 Gyr in a series of major mergers, 
the physical mass distribution [not $M_{\rm vir}(z)$] remains practically constant 
since $z=1$. The same is true in a large sample of $\Lambda$CDM galaxy halos.
Subhalo mass loss peaks between
the turnaround and virialization epochs of a given mass shell,
and declines afterwards.  
97\% of the $z=1$ subhalos have a surviving bound remnant 
at the present epoch. The retained mass fraction is 
larger for initially lighter subhalos: satellites with maximum circular 
velocities $\vmax=10\,\kms$ 
at $z=1$ have today about 40\% of their mass back then. At the first pericenter
passage a larger average mass fraction is lost than during each following orbit.
Tides remove 
mass in substructure from the outside in, leading to higher concentrations compared to 
field halos of the same mass. This 
effect, combined with the earlier formation epoch of the inner satellites, results in strongly 
increasing subhalo concentrations towards the Galactic center. We present individual 
evolutionary tracks and present-day properties of the likely hosts of the dwarf 
satellites around the Milky Way. The formation histories of ``field 
halos'' that lie today beyond the Via Lactea host are found to strongly depend on the density 
of their environment. This is caused by tidal mass loss that affects many field halos on 
eccentric orbits.
\end{abstract}

 
\section{Introduction}
Cosmological N-body simulations with large dynamic range (i.e.\ with large
numbers of particles per virialized object and adequately high force
and time resolution) make it possible to follow the highly non-linear
formation of cold dark matter (CDM) halos and their substructure in
great detail \citep[e.g.][]{Ghigna1998,Ghigna2000,Klypin1999,Moore1999,Moore2001,
Fukushige2004,Kravtsov2004,Diemand2004sub,Gao2004,Gill2005,Reed2005}.
We have recently completed ``Via Lactea'', the highest resolution
simulation to date of CDM substructure. The run was completed
in 320,000 CPU hours on NASA's Project Columbia supercomputer, and
follows the formation of a Milky Way-size halo with 234 million
particles, an order of magnitude more than achieved previously.  The
present-day properties of the galaxy host and its substructure were
presented in Diemand, Kuhlen, \& Madau (2007, hereafter Paper I). In
this second paper we use data extracted from all 200 snapshots stored
during the Via Lactea run to study the mass assembly history of the
main halo and the subhalo population.

\citet{Ghigna1998} and \citet{Bullock2001conc} have noted that subhalos are
more concentrated than field halos. We now have the resolution and statistics
to quantify this effect, and the large number of snapshots allows us to understand its
origin. We can follow the evolution of
massive subhalos at similar or higher resolution
as in idealized N-body experiments that evolve one satellite in an
external potential \citep[e.g.][]{Hayashi2003,Dekel2003,Kazantzidis2004,Read2006}, but
within a ``live'' host halo forming and evolving within 
the cosmological context. The increased
resolution allows more accurate estimates of the fraction of subhalos
that survive, the mass they retain, the effect of tidal stripping on
their internal structure. How subhalo density profile and
concentrations evolve during tidal mass loss?
Are subhalos really fully disrupted once the tidal radius
at pericenter is smaller than their scale radius \citep{Hayashi2003}? 
How strongly does the heating from tidal shocks reduce inner
subhalo densities? Or do tides in the inner, shallow part of the host
lead to subhalo compression \citep{Dekel2003}?
 
One interesting result that becomes apparent in cosmological simulations when tracking
halos moving within a cluster potential is that many (sub)halos that 
were well within the cluster 
virial radius $\rvir$ at some earlier time can be found today beyond $\rvir$
\citep{Balogh2000,Moore2004,Gill2005}. This implies that the formation histories of ``field'' galaxy 
halos are affected by their environment, as many of the halos found in the outskirts of 
larger systems today may have shed mass during an earlier pericenter 
passage because of tidal interactions. Significant correlations between the formation times of 
galaxy-size halos of a fixed mass, their clustering strength \citep{Sheth2004,Gao2005}, and 
the density of their environment \citep{Harker2006} have indeed been found. In this 
work, we show that these correlations are caused by tidal interactions of (sub)halos on 
extended radial orbits with a more massive neighbor. 

This paper is organized as follows.
In \S~\ref{section:formation} we study the mass assembly of the Via Lactea host and its
substructure population, and introduce physical (non-comoving) general definitions of (sub)halo 
properties like size, mass, concentration, and formation time. Section \ref{section:fixedmass} 
discusses the evolution of subhalo concentrations and abundance in fixed-mass shells 
around the host. In \S~\ref{tracks} we analyze individual and ensemble-averaged evolutionary 
tracks of subhalos and discuss how their density profiles evolve during tidal shocks at 
pericenter. We present the histories and present-day properties of the likely hosts of the 
dwarf satellites around the Milky Way. Average histories of subhalos found in certain regions 
today and their survival from $z=1$ to $z=0$ are also discussed in this section.
Finally, \S~\ref{section:concl} summarizes our conclusions.

\section{Formation histories of galaxy halos}
\label{section:formation}

\begin{figure}
\epsscale{1.2}
\plotone{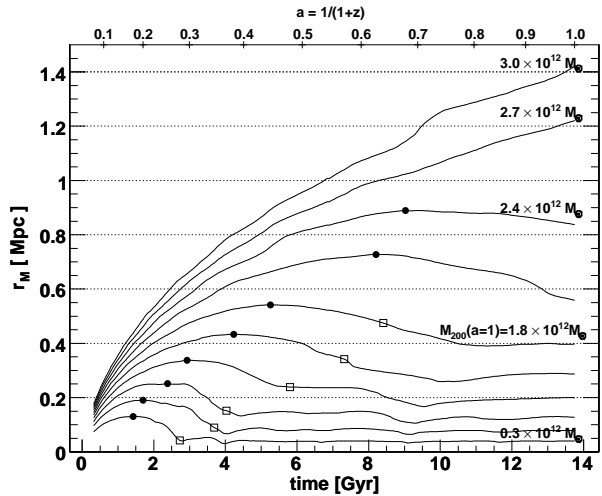}
\caption{Evolution of radii $r_M$ enclosing a fixed mass versus cosmic time or scale factor $a$. 
The enclosed mass grows in constant amounts of $0.3\times 10^{12}\,\msun$
from bottom to top. Shells are numbered from one (inner) to ten (outer).
Initially all spheres are growing in the physical 
(non comoving) units used here. Shells 1 to 6 turn around, collapse and stabilize, while
the outermost shells are still expanding today. {\it Solid circles}: points of 
maximum expansion at the turnaround time $t_{\rm ta}$. {\it Open squares:} time after 
turnaround where $\rfm$ first contracts within $20\%$ of the final value. These mark 
the approximate epoch of stabilization. The collapse factors $r_M(t_{\rm ta})/r_M(z=0)$ for 
shells 1 to 6 are 3.29, 2.44, 1.98, 1.70, 1.51 and 1.36, respectively. Thus shells 1 and 
2 collapse by more than the factor of 2 derived from spherical top-hat, while shells 4, 5,
and 6 collapse by a smaller factor.
}
\label{LagRadii}
\end{figure}

The Via Lactea simulation was performed with the PKDGRAV tree-code
\citep{Stadel2001} and employed multiple mass particle
grid initial conditions generated with the GRAFICS2 package
\citep{Bertschinger2001}. The high resolution region was sampled with
234 million particles particles of mass $2.1\times 10^4\msun$ and evolved
with a force resolution of 90 pc. It was embedded within a periodic
box of comoving size 90 Mpc, which was sampled at lower resolution to
account for the large scale tidal forces. We adopted the best-fit
cosmological parameters from the {\it WMAP} 3-year data release
\citep{Spergel2006}: $\Omega_M = 0.238$, $\Omega_\Lambda = 0.762$,
$H_0= 73\,\kmsmpc$, $n=0.951$, and $\sigma_8=0.74$.
The simulation was centered on an isolated halo that had no major
merger after $z=1.7$, which makes it plausible that this halo would 
be a suitable host for a Milky Way-like disk galaxy \citep[e.g.][]{Fabio2007}.
More details about the Via Lactea run are
given in Paper I. Movies, images, and data are 
available at http://www.ucolick.org/$\sim$diemand/vl. The host halo mass at 
$z=0$ is $\mtwo=1.77\times 10^{12}\,\msun$ within a radius of $\rtwo=389\,$ 
kpc (we define $\rtwo$ as the radius within which the enclosed average 
density is 200 times the mean matter density $\Omega_M\,\rho_{\rm crit}$. Note that 
$\mtwo$ and $\rtwo$ were denoted in Paper I as $\rvir$ and $\mhalo$. We revert 
here to the more standard notation for reasons of clarity.).

\begin{figure}
\epsscale{1.2}
\plotone{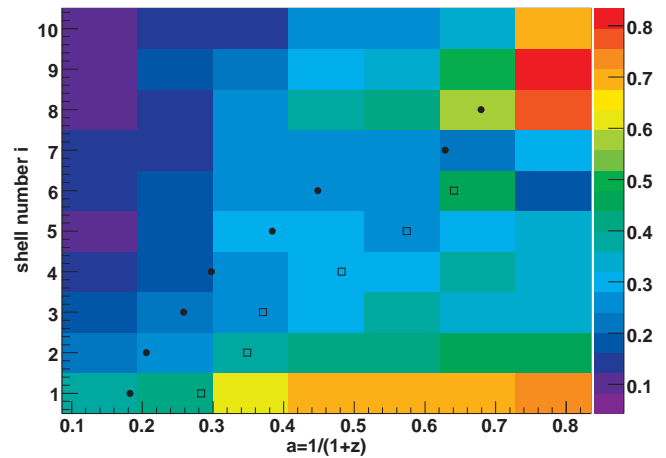}
\caption{Fraction of material belonging to shell $i$ at epoch
$a$ that remains in the same shell today. Shells are same as in 
Fig. \ref{LagRadii}, numbered from one (inner) to ten (outer).
{\it Solid circles}: time of maximum expansion. {\it Open squares:} 
stabilization epoch. Mass mixing generally decreases with time and towards 
the halo center.
}
\label{overlap}
\end{figure}

\begin{figure*}
\plotone{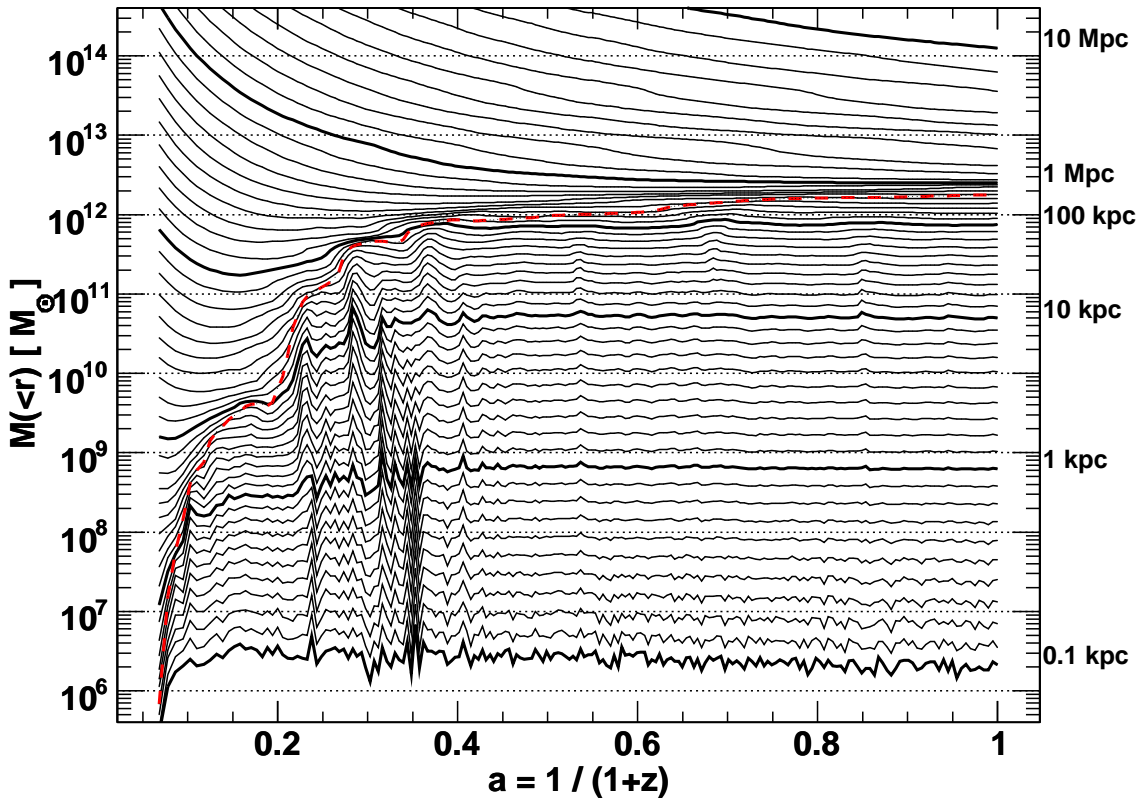}
\caption{Mass accretion history of Via Lactea. Masses within 
spheres of fixed physical radii centered on the main progenitor are plotted
against the cosmological expansion factor $a$. The thick solid lines correspond
to spheres with radii given by the labels on the right. The thin solid lines 
correspond to nine spheres of intermediate radii that are 1.3, 1.6, 2.0, 2.5, 3.2, 4.0, 
5.0, 6.3 and 7.9 times larger than the next smaller labeled radius. {\it Dashed line:} 
$M_{200}$. The halo is
assembled during a phase of active merging before $a\simeq0.37$ ($z\simeq1.7$) and
remains practically stationary at later times.}
\label{massaccr_norvcm}
\end{figure*}

Before describing the evolution of substructure, we have to
address the following issues. When does a halo become a subhalo?
When does the host form and how does it grow? Which regions/volumes 
should be used for a meaningful comparison of subhalo abundances and 
average properties at different cosmic epochs? The common procedure is 
to define at each epoch a ``virial" radius $\rvir$, which depends 
on the cosmic background density at the time, and define subhalos as bound clumps 
within this volume. These definitions are not ideal
for two reasons. First, halos cross this artificial boundary not only inward
(``accretion") but about as often also outwards. Averaging over six, relaxed galaxy 
clusters, no net infall of subhalos into the virial region was found in 
\citet{Diemand2004sub}, and half of the halos found today between $\rvir$ and $2\,\rvir$ 
had actually passed through the cluster at some earlier time \citep{Balogh2000,Moore2004,Gill2005}. 
\citet{Maccio2003} noted that the virialized regions of halos are often larger than 
$\rvir$, and a lack of mass infall out to 2-3 virial radii was found for a large, 
representative sample of galaxy halos by \citet{Prada2006}. This leads directly to the
second problem. As the cosmic background density decreases with Hubble expansion, formal 
virial radii and masses grow with cosmic time even for stationary halos. Studying 
the transformation of halo properties within $\rvir$ (or some fraction of it) mixes 
real physical change with apparent evolutionary effects caused by the growing radial 
window, and makes it hard to disentangle between the two.

To address the second problem, we describe here the formation of Via Lactea 
using radial shells enclosing a fixed mass, $\rfm$. Unlike $\rvir$, $\rfm$ stops
growing as soon as the mass distribution of the host halo becomes stationary on the 
corresponding scale (see Fig. \ref{LagRadii}). The first problem, however, remains. 
Mass and substructure are constantly exchanged between these shells, as $\rfm$ is 
not a Lagrangian radius enclosing the same material at all times, just the same
amount of it. The fraction of material belonging to a given shell in the past 
that still remains within the same shell today 
is shown in Figure \ref{overlap}. The mixing is larger before 
stabilization, presumably because of shell crossing during collapse. In the stationary 
phase the shells still exchange mass because many particles are on radial orbits. 
The mixing is smaller near the halo center, where most of the mass is in a
dynamically cold, concentrated component that originated from the earliest 
forming high-$\sigma$ progenitors \citep{Diemand2005}.

\subsection{Collapse times and collapse factors}
\label{collapse}

In spherical top-hat collapse, a shell has no kinetic energy at turnaround
and virializes at half the turnaround radius. The final overdensity
relative to the critical density at the collapse redshift is
$\Delta=18\pi^2$ in the Einstein-de Sitter model, modified
in a flat Universe with a cosmological constant to the fitting formula
(Bryan \& Norman 1998)
\begin{equation}
\Delta=18\pi^2-82\Omega_\Lambda(z)-39\Omega_\Lambda^2(z),
\end{equation}
where
\begin{equation}
\Omega_\Lambda(z)=\frac{\Omega_\Lambda}{\Omega_m(1+z)^3+\Omega_\Lambda}.
\end{equation}
At $z=0$ and for a {\it WMAP} 3-year cosmology, this yields $\Delta=93$.
Here we introduce the modified formula,
\begin{equation}
\Delta=200-82\Omega_\Lambda(z)-39\Omega_\Lambda^2(z),
\end{equation}
and define the virial radius $\rvir$ as the radius enclosing a mean
density $\Delta\rho_{\rm crit}$. At $z=0$ this yields $\Delta=104$
and $\rvir = 288$ kpc for Via Lactea. We chose this slightly
different definition for the collapse overdensity so that, at high
redshifts, $\rvir$ approaches $\rtwo$.

The simple spherical top-hat collapse ignores shell crossing and
mixing, triaxiality, angular momentum, random velocities, and large
scale tidal forces.  Figure \ref{LagRadii} shows that spheres
enclosing a fixed mass have collapse factors that differ from 2. Inner
shells collapse by larger factors, in qualitative agreement with the
modified spherical collapse model of \citet{Sanchez2006} that accounts
for shell crossing but not angular momentum. Shells enclosing about
the standard viral mass collapse by less than a factor of 2, probably
because of the significant kinetic energy they contain already at
turnaround. The collapse times are also different from spherical
top-hat. Shell number five, for example, encloses a mean density of
about $104\,\rho_{\rm crit}$ today, a virial mass of
$1.5\times10^{12}\,\msun$ and should have virialized just now
according to spherical top-hat. It did so instead much earlier, at
$a=0.6$. Even the next larger shell with $1.8\times10^{12} \msun$
stabilized before $a=0.8$. Our analysis supports the point made by
\citet{Prada2006}, that spherical top-hat provides only a crude
approximation to the virialized regions of simulated galaxy halos.

\begin{figure*}
\plotone{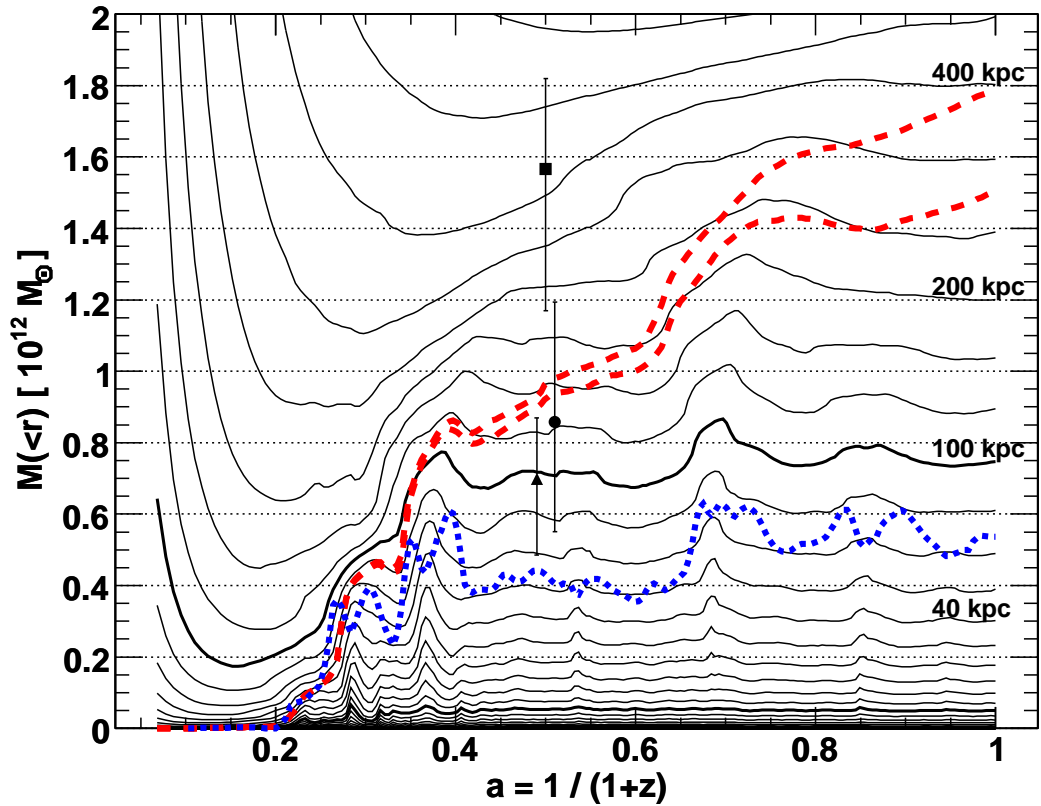}
\caption{Same as Figure \ref{massaccr_norvcm}, but using a linear scale in enclosed mass.
In addition to $M_{200}$ ({\it upper dashed line}) we now also plot $M_{\rm vir} $
({\it lower dashed line}) and the mass within the radius of 
maximal circular velocity ({\it dotted line}). The physical mass accretion is small after
the last major merger at $a\simeq0.37$ ($z\simeq1.7$): more than 80\% of the present-day 
material within 400 kpc is already in place at $z=1$. This value is typical for
galaxy-size halos. {\it Filled square:} median $z=1$ mass fraction ($=0.87$) within
400 kpc for 303 halos of similar mass rescaled to today's Via Lactea mass within 400 kpc.
{\it Solid circle}: corresponding median $z=1$ value for $M_{200}$. {\it Filled triangle:} 
corresponding median $z=1$ value for the mass within 100 kpc. Error bars indicate the 68\%
scatter around the median.}
\label{massaccrlin}
\end{figure*}

\vspace*{0.1in}
\subsection{Accretion histories}

To understand the mass accretion history of the Via Lactea halo 
we now analyze the evolution of mass within fixed physical radii.
Figure \ref{massaccr_norvcm} shows that the mass within all radii from 
the resolution limit of $\simeq $1 kpc up to 100 kpc grows during a series 
of major mergers before $a=0.4$.  After this phase of active merging and
mass accretion the entire system is almost perfectly stationary at all
radii. The small decrease in density on scales below 1 kpc is likely
an artifact of time-steps which are too large compared to the short
dynamical time in these inner regions (Paper I). 
A similar decrease in density in the inner regions of halos was shown to
be caused by too large time steps in the convergence tests
of Fukushige et al (2004; see their Figure 9).
Only the outer regions ($\sim $ 400 kpc) experience a small amount of
mass accretion after the last major merger, better visible in the
linear mass scale of Figure \ref{massaccrlin}.  The mass within 400
kpc increases only mildly, by a factor of 1.2 from $z=1$ to the present.
During the same time the mass within radii of 100 kpc and smaller,
the peak circular velocity, $\vmax$, and the radius where it is reached, $\rvmax$, 
all remain constant to within 10\%.
The lack of evolution in the inner density profile, and therefore also in 
$\vmax$, $\rvmax$ and $\rho(<\rvmax)$, during this major merger-free phase
agrees with the findings of previous studies
\citep[e.g.][]{Wechsler2002,Zhao2003,Romano2006}.

The physical assembly of galaxy halos thus appears to occur mainly
during an active early phase of major mergers, in which the halo peak
circular velocity (and the enclosed mass at all radii) grows to its
maximum value. In contrast to previous work on this subject
\citep{Wechsler2002,Zhao2003}, we do not find evidence for much mass
growth during the late ``slow accretion'' phase, \textit{when the definition of halo
mass is based on physical instead of comoving scales}. Rather, the 
mass distribution and peak circular velocity appear to remain constant for the
majority of the halo's lifetime.
The fact that mass definitions inspired by spherical top-hat 
fail to accurately describe the real assembly of galaxy halos is clearly seen in 
Figure \ref{massaccrlin}, where $M_{200}$ and $M_{\rm vir}$ are shown to
increase even when the halo physical mass remains the same.
This is just an artificial effect caused by the growing radial windows 
$\rvir$ and $\rtwo$ as the background density decreases. For Via Lactea $M_{200}$ 
increases by a factor of 1.8 from $z=1$ to the present, while the real 
physical mass within a 400 kpc sphere grows by only a factor of 1.2 during the 
same time interval, and by an even smaller factor at smaller radii.

We find that the small physical accretion since $z=1$ seen in Via Lactea 
is indeed typical of galaxy halos. In a 45 Mpc periodic box resolved with $300^3$
particles of mass $1.2\times10^8 \msun$ (simulated in a 3-year {\it WMAP}
cosmology), we have identified 303 galaxy halos at $z=0$ with $M_{200}$
ranging from $0.6\times10^{12} \msun$ to $5.4\times10^{12} \msun$.
The mass within a constant physical radius of 400 kpc grows by a
factor of $1.15_{-0.16}^{+0.39}$ (the errors indicate the 68\% range
around the median) since $z=1$ (``physical accretion''), whereas
$M_{200}$ grows by a factor of $2.10_{-0.59}^{+1.17}$ (``apparent accretion''). 
For lower mass halos the physical
accretion is even smaller. In the same 45 Mpc box we find 714 halos with 
$M_{200}$ ranging from $1.5\times10^{11} \msun$ to $4.5\times10^{11} \msun$
today. From $z=1$ to the present their mass within 200 kpc ($\sim r_{200}$ at $z=0$)
grows by a factor of only $1.12_{-0.17}^{+0.26}$, whereas $M_{200}$ increases by 
$1.85_{-0.40}^{+0.96}$. Physical and apparent accretion are correlated, 
but with a large scatter.

Within the inner 20 kpc, i.e.\ where the galaxy is expected to lie, the gravitational
potential remains constant during the late, quiescent phases of halo formation (Figure
\ref{massaccr_norvcm}). Unless there is an evolving, dominant baryonic
mass contribution, the rotation rate of the galactic disk should not
evolve in time, with a peak circular velocity that may be proportional
to the constant peak circular velocity of its halo. Assumptions
sometimes made in semi-analytic models about evolution of galaxy
properties with halo virial quantities (e.g. stellar mass with
$\mvir$) will produce inaccurate results, considering
the different length scales and the lack of physical accretion both on
small and large scales. Models based on quantities that do remain constant 
during stationary phases, like peak circular
velocity and the corresponding radius and enclosed mass, may be more physical.
Observations of representative samples of $z=1$ galaxies, including kinematics, are
now becoming available \citep[e.g.][]{Weiner2006,Kassin2007} and it might be
possible to test whether galaxy radii and masses grow like the halo
virial scales (i.e.\ by about a factor of two) or if they remain
constant like the halo mass distribution on physical scales.

\subsection{Formation times}

As discussed above, the common spherical top-hat inspired halo mass
definitions $M_{200}$ and $M_{\rm vir}$ are not well suited to
describe the growth of galaxy halos. Care should also be used when the
complex and extended process of halo formation is quantified with a
single number, the so called ``halo formation time''. Many of the
existing definitions are based on the evolution of $M_{200}$ or
$M_{\rm vir}$ \citep[e.g.][]{Wechsler2002,Zhao2003,Gao2005}.  For galaxy halos
such formation times depend almost exclusively on the amount of apparent
accretion
\footnote{The amount of apparent accretion depends on how much mass lies in the outer halo,
i.e.\ it is larger for halos with low concentrations.}
, which dominates the evolution of $M_{200}$ and $M_{\rm vir}$ 
for more than half of the age of the universe and, in many cases, contributes
more than half of the total ``accreted'' mass. Since apparent accretion correlates
only weakly with physical halo growth, it is unclear how and if halo formation 
times calculated in this manner do relate to the epoch when most
of the physical halo assembly took place.
Our analysis also casts doubt on whether such formation times are at all
related to the relevant timescales for galaxy formation.

Consider, as an example, the widely used definition
of formation time as the time when $M_{200}$ (or similarly the
FOF mass based on a comoving linking length of 0.2 times the mean
particle separations) reaches half of the present value. More than
half of our large, low-resolution sample of galaxy-size halos would
form after $z=1$ according to this definition, yet their physical mass
accretion is less than 20\% over this time span; \textit{their mass
assembly was practically completed before their formal ``formation
time''}. The Via Lactea halo would have a formation redshift of $z\simeq1$
according to this definition (see Figure \ref{massaccrlin}), which is
also well after the epoch when most of the physical mass accretion 
actually took place.
To address this issue, we propose a formation time based on peak circular velocity, a
quantity that does not evolve during the stationary phase of a 
halo. We define the halo formation epoch $z_{\rm form}$ to be equal to
the earliest time when $\vmax$ reaches 85\% of its highest value at all redshifts:
\be \label{zform}
V_{\rm max} (z_{\rm form}) \equiv 0.85 \;  \mathop{\rm max}_{z} \{ V_{\rm max} (z) \} \;. 
\ee
Note that a definition of formation time based on the present-day peak circular
velocity $V_{\rm max}(z=0)$ would lead to significantly higher median
formation redshifts, since for many halos $V_{\rm max}$ is
reduced by tidal stripping. We will show in Section
\ref{section:envdep} that this is true even for halos beyond the
virial radius today, i.e.\ for ``field'' halos.  For comparison, we
define the redshift $z_{\rm 85}$ as
\be \label{z85}
V_{\rm max} (z_{\rm 85}) \equiv 0.85 \;  V_{\rm max} (z=0) \; ,
\ee
but throughout this work we mean $z_{\rm form}$ (eq. \ref{zform})
when we refer to a halo formation time.  In Section
\ref{section:envdep} we will find a clear environmental dependence in the
median $z_{\rm 85}$ of field halos, but not in their median $z_{\rm
form}$.

\subsection{A physical (sub)halo concentration index: $c_V$}

Often halo concentrations are presented in terms of the virial
concentration index defined as the ratio $\cvir = \rvir / r_s$, where
$r_s$ is the scale radius of an NFW fit
\citep[e.g.][]{Navarro1997,Bullock2001conc,Wechsler2002,Kuhlen2005,Maccio2006}.
This definition has two drawbacks: 1) $\cvir$ grows even during epochs
of ``apparent accretion'', when the physical mass distribution remains
constant. In Via Lactea, for example, the mass distribution remains
nearly unchanged from $z=1$ to $z=0$, but $\cvir$ grows by about a
factor of two because of the comoving definition of $r_{\rm vir}$; and 2) subhalos 
are truncated at the tidal radius, which is always smaller
than their formal virial radius, i.e.\ virial radii and thus $\cvir$ are not well 
defined for subhalos.

A direct measure of physical density in the inner regions of halos is
provided by the ``central density parameter'' $\Delta_{V/2}$,
introduced by \citet{Alam2002}. Here we refer to this parameter as
$c_V$, to avoid confusion with the virial overdensity $\Delta$.
\be
c_{V/2} \equiv \f{\bar{\rho}(<r_{\rm Vmax/2})}{\rho_{\rm crit,0}} = \f{1}{2} \left( \f{V_{\rm max}}{H_0 r_{\rm Vmax/2}} \right)^2,
\ee
where $r_{\rm Vmax/2}$ is the radius at which the circular velocity
curve reaches half its maximum value. With this definition $c_{\rm V/2}
\rho_{\rm crit,0}$ is equal to
the mean physical density within $r_{\rm Vmax/2}$ and has the
advantage that it can be directly measured in numerically simulated
dark matter halos and in observed galactic rotation curves,
\textit{without reference to any particular analytic density
profile}. Unfortunately, even with Via Lactea's extreme resolution, an
accurate determination of $r_{\rm Vmax/2}$ is not possible for all but
the most massive subhalos. Since $r_{\rm Vmax}$ is better measured, however, 
we use instead $c_V$, the mean physical
density within the radius of the peak circular velocity in units
of $\rho_{\rm crit,0}$, as our new physical concentration parameter:
\be \label{eq:conc}
c_{V} \equiv \f{\bar{\rho}(<r_{\rm Vmax})}{\rho_{\rm crit,0}} = 2 
\left( \f{V_{\rm max}}{H_0 r_{\rm Vmax}} \right)^2.
\ee
\begin{figure}
\epsscale{1.2}
\plotone{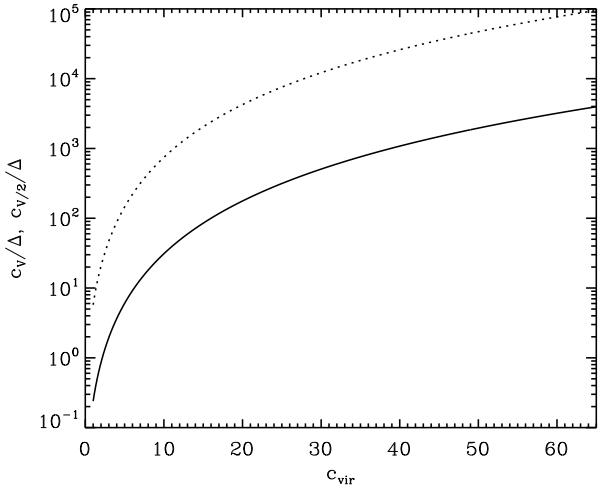}
\caption{
Concentration parameters $c_V$ (\textit{solid}) and
$c_{V/2}$ (\textit{dotted}) divided by the density contrast
$\Delta$ used to define $\rvir$ at $z=0$, as a function of 
$c_{\rm vir} = \rvir/r_s$.}
\label{fig:DeltaV}
\vspace*{0.1in}
\end{figure}
For any given analytic density profile it is straightforward to
convert between $\cvir$ and $c_V$. For an NFW
\citep{Navarro1997} density profile,
\be
\rho(r) = \frac{\rho_s}{r/r_s (1 + r/r_s)^2},
\ee
the circular velocity is 
\ba
V_c(r) & = & 4 \pi G \rho_s r_s^3 \f{f(r)}{r}, \quad {\it with} \label{eq:vcirc} \\
f(r) & = & \ln(1 + r/r_s) - \frac{r/r_s}{1 + r/r_s}.
\ea
The maximum of $V_c(r)$ occurs at
\be
r_{\rm Vmax} = 2.163 \; r_s.
\label{eq:rvmax}
\ee
The NFW scale density $\rho_s$ can be expressed in terms of the
concentration $\cvir$ and the spherical top-hat virial density
contrast $\Delta$ (cf. Section~\ref{collapse}) 
\be
\rho_s = \f{1}{3} \f{\cvir^3}{f(\rvir)} \Delta(z) \rho_{\rm crit}(z). \label{eq:rhos}
\ee
Combining eqs. (\ref{eq:vcirc}), (\ref{eq:rvmax}), and (\ref{eq:rhos}), we find 
\be
c_V =\left(\f{\cvir}{2.163}\right)^3  \f{f(r_{\rm Vmax})}{f(\rvir)} \Delta(z) \; \frac{\rho_{\rm crit}(z)}{\rho_{\rm crit,0}} .
\ee

Figure~\ref{fig:DeltaV} shows a plot of $c_V$ divided by the density contrast
$\Delta$ used to define $\rvir$ and $\cvir$, as a function of $\cvir$.
For comparison, we also show $c_{V/2} / \Delta$. Note that
$c_V$ is defined in terms of $\rho_{\rm crit}$ today,
whereas $\cvir$ explicitely depends on redshift
through $\Delta(z)$ and $\rho_{\rm crit}(z)$. The values of $c_V$ for a given $\cvir$ that can
be read off from Figure~\ref{fig:DeltaV} are thus only valid at $z=0$;
at higher redshifts they must be multiplied by 
$[\Delta(z)\rho_{\rm crit}(z) ] / [\Delta(0)\rho_{\rm crit,0}]$. 
At $z=0$ the Via Lactea host halo has a concentration of
$c_V = 3613$, which corresponds to $\cvir=10.4$. This
compares well with the value of $\cvir = 11.7$ determined from
the best fitting NFW model.

\begin{figure*}
\plotone{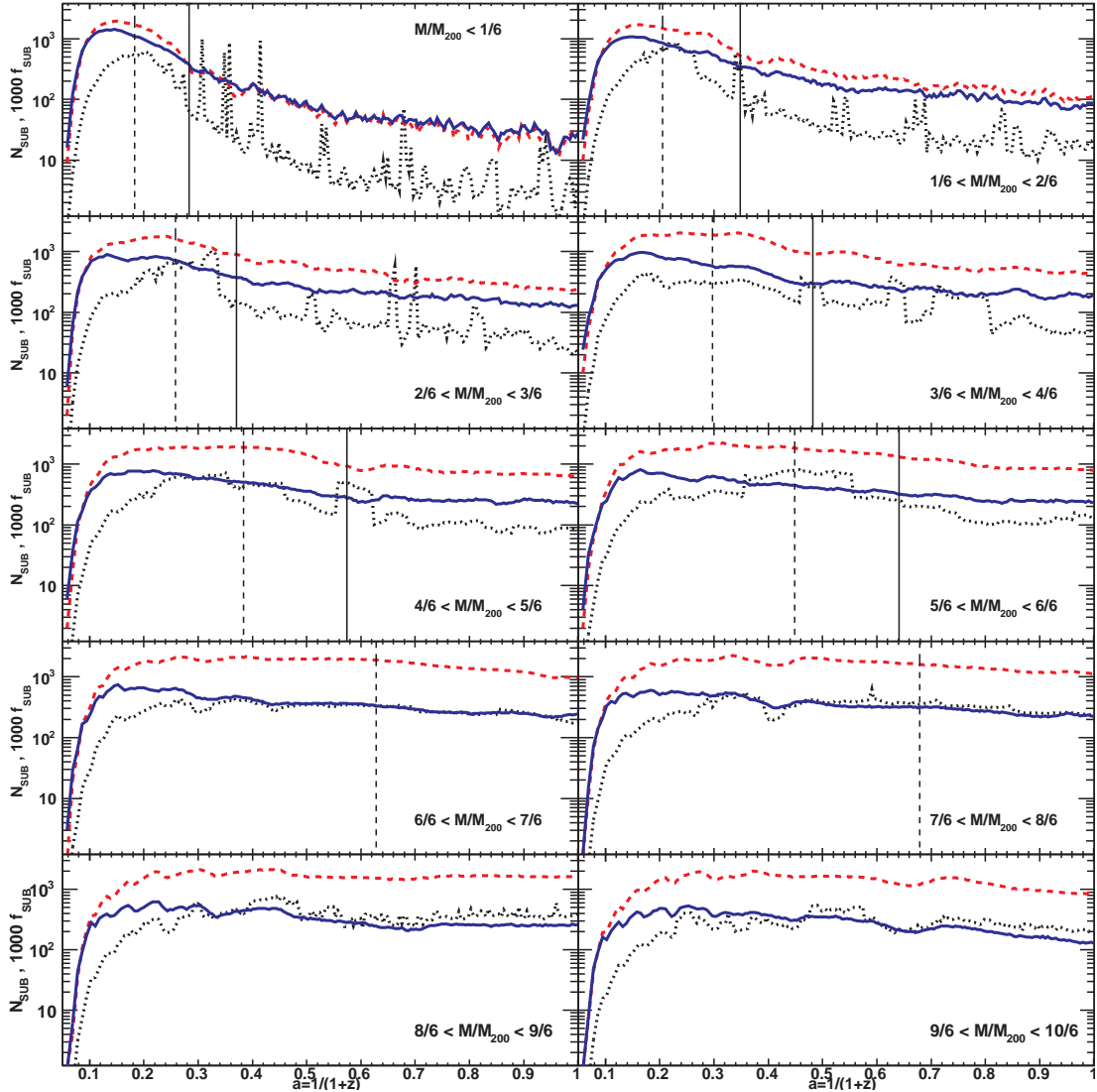}
\caption{Abundance of (sub)halos versus time in shells containing a fixed mass and
centered on the main Via Lactea progenitor at each time. Shells are ordered
from inner ({\it top left}) to outer ({\it bottom right}). {\it Solid line:} number
of subhalos with $\vmax > 5\,\kms$.{\it Dashed line:} number of subhalos
with $\msub > 4.0\times10^6 \msun$.{\it Dotted line:} mass
fraction in resolved (sub)halos within each shell (excluding the most
massive subhalo to avoid spikes as it orbits though the shells). The vertical dashed 
line marks the time of maximal expansion of the corresponding mass shell, and the
vertical solid line the approximate stabilization epoch (see Figure \ref{LagRadii}).
The subhalo mass loss rate peaks between these two epochs and
declines after the region stabilizes.
}
\label{massfrEv}
\end{figure*}

\section{Evolution of subhalo properties}
\label{section:fixedmass}

The present-day cumulative subhalo mass function within Via Lactea 
is well approximated by a simple power law
\begin{equation}
N(> \msub) = 0.0064\; \left( \frac{\msub}{M_{200}}\right)^{-\alpha_M} \; ,
\end{equation}
with slope\footnote{$N(> \msub)$ is not a perfect power law: It becomes steeper at large
masses, due to dynamical friction, and shallower at small masses, due to the
gradually increasing importance of numerical resolution effects.
Therefore the best fit slope depends on the mass range and the fitting procedure.
For $\msub>200 \, m_p$ we find $\alpha_M = 0.97 \pm 0.03$. In the same mass range
the differential mass function $dn(\msub)/d\msub$ has
$\alpha_{dM} = 1.90 \pm 0.02$. These best fit values differ by less than unity, because
the differential mass function gives more weight to the relatively poorly resolved low mass end.}
$\alpha_{M} \simeq 1$ and host halo mass $M_{200} = 1.8
\times 10^{12} \msun$ (Paper I). Here $\msub$ is defined as the mass within the
tidal radius $r_t\equiv r \sigma_{\rm sub}/(\sqrt{2}\sigma_{\rm host})$. This radius
is the classical Jacobi limit for an isothermal satellite on a circular orbit of radius r within an
isothermal host halo (see e.g. \citealt{Read2006}). It has the property that the host local
density is half of the local satellite density at $r_t$. Similarly the $z=0$
subhalo velocity function within $r_{200}$ is fitted by
\begin{equation}
N(>V_{\rm max})= 0.021 \left( \f{V_{\rm max}}{V_{\rm max,host}}\right)^{-\alpha_V}  \; ,
\end{equation}
with slope $\alpha_{V} \simeq 3$ and $V_{\rm max,host}=182 \kms$. Here we present the time evolution
of the normalizations and slopes of these two power laws. 

\begin{figure*}
\epsscale{1.2}
\plotone{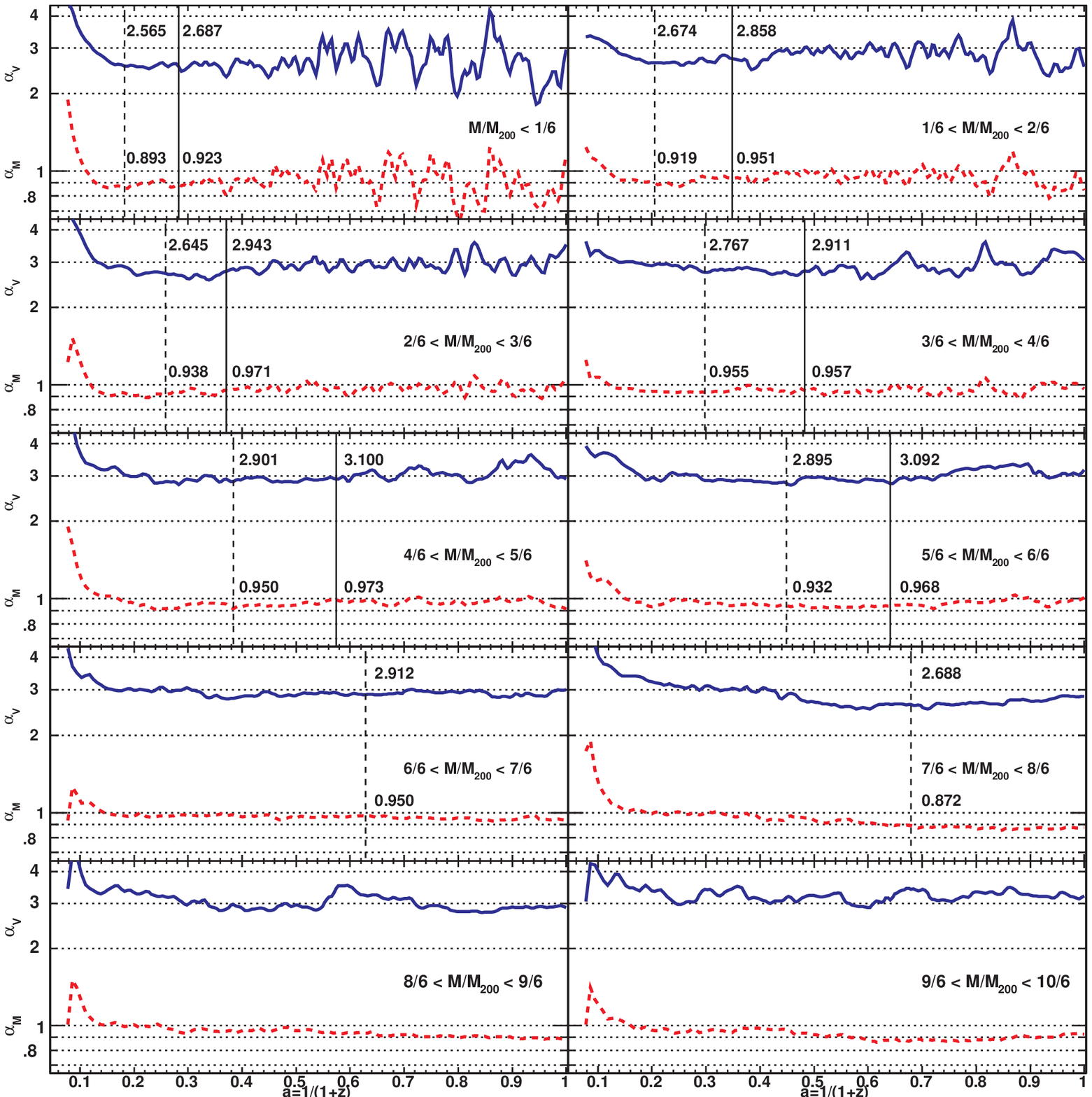}
\caption{Evolution of the slopes of the cumulative subhalo velocity ({\it solid lines}) 
and mass ({\it dashed lines}) function in the same shells as in Fig. \ref{massfrEv}.
Numbers depict the average slopes between the turnaround and stabilization epochs, and 
from stabilization to the present. The slopes show little trend with time or distance 
from the main progenitor. 
}
 \label{slopeEv}
 \end{figure*}

\begin{figure*}
\epsscale{1.2}
\plotone{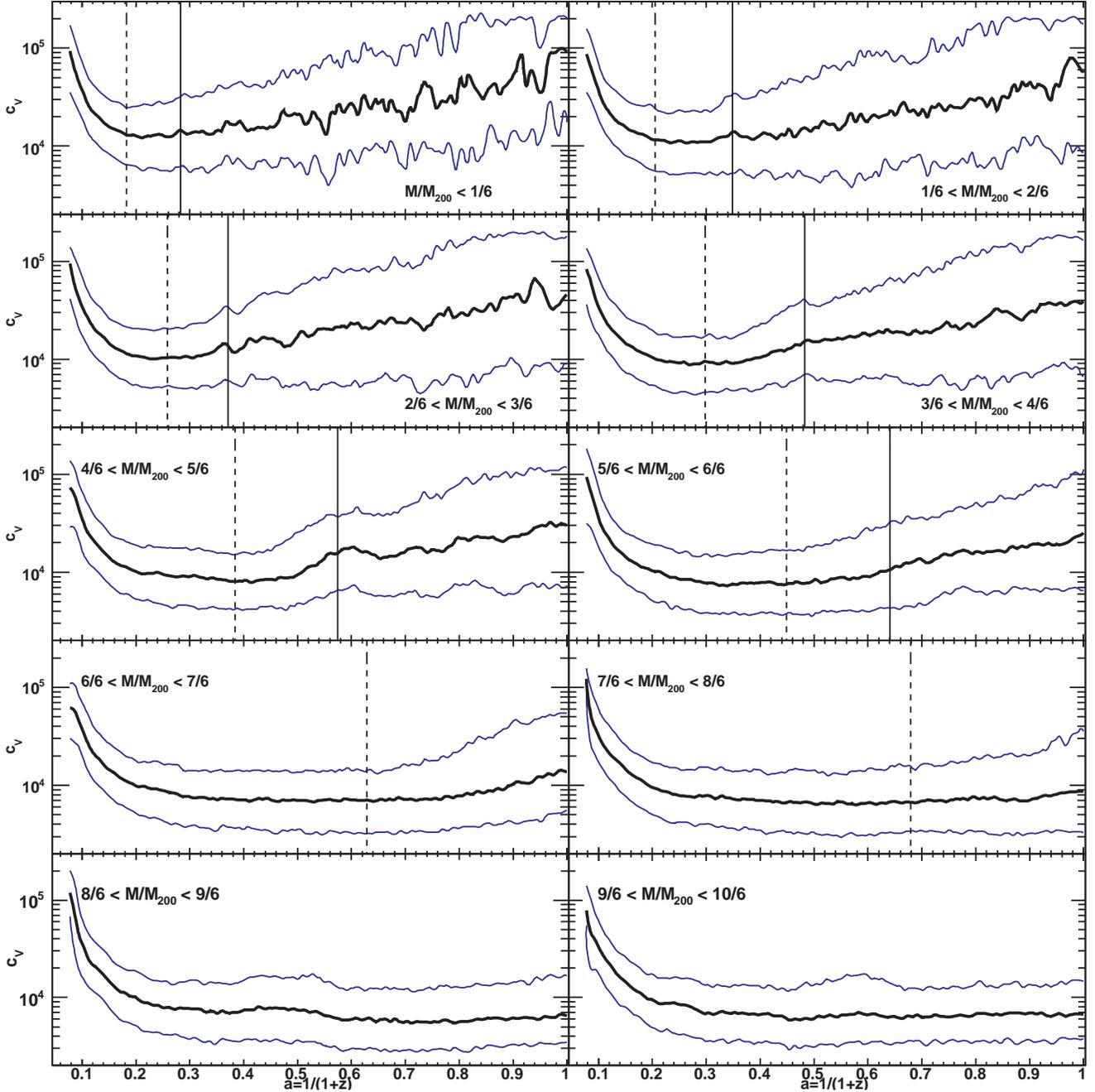}
\caption{Same as Fig. \ref{massfrEv}, but now the evolution of the median subhalo 
concentration ({\it thick line}) is plotted versus scale factor. {\it Thin line:} 
$68\%$ scatter around the median. All halos with $\vmax > 5\,\kms$ are included.
}
\label{concEv}
\end{figure*}

\begin{figure}
\epsscale{1.2}
\plotone{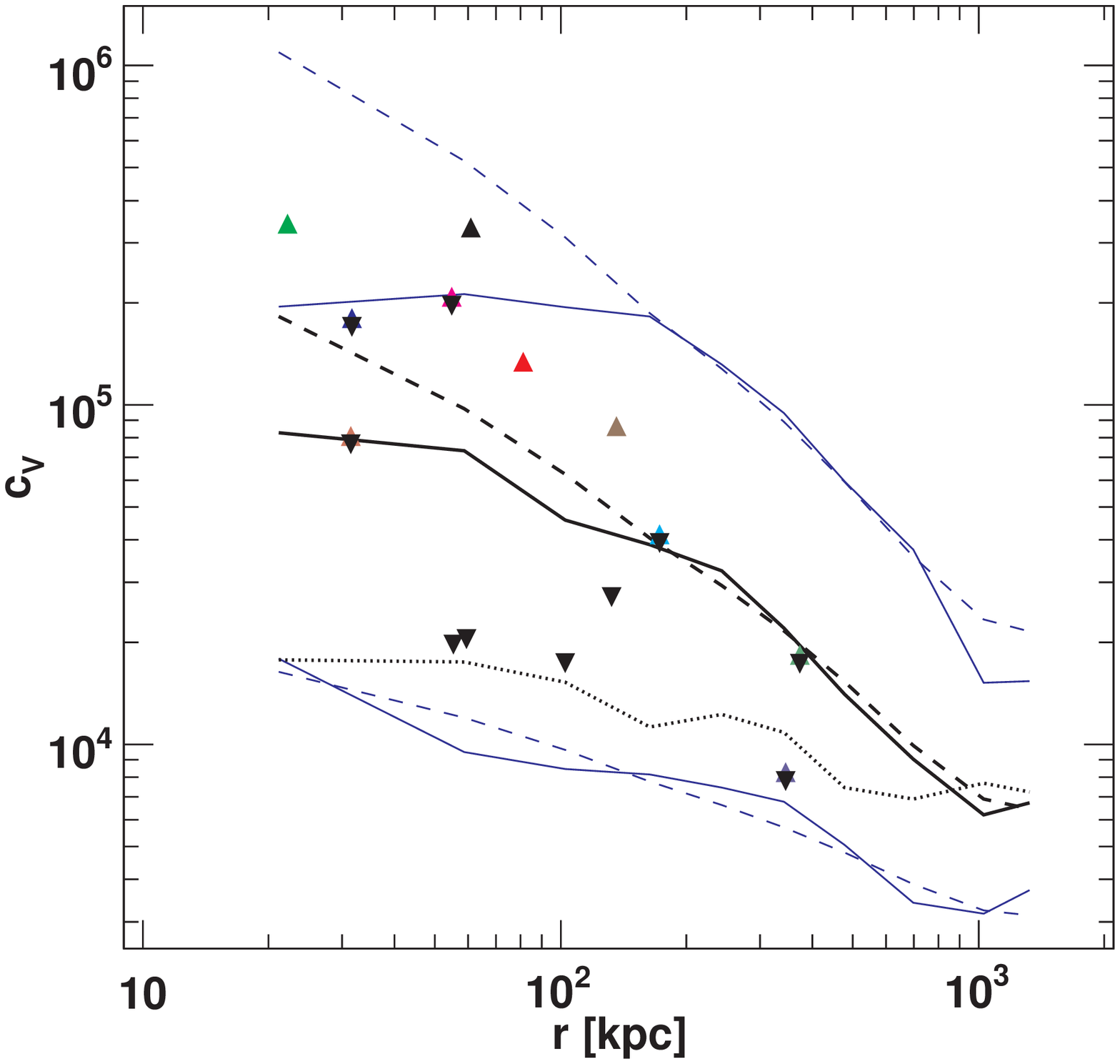}
\caption{Median subhalo concentrations and $68\%$ scatter ({\it solid lines})
versus distance from the Galactic center at the present epoch (average values
over the last ten snapshots from $z=0.05$ to $z=0$). 
The dotted line shows 400 times the cosmic background density at the median formation
times of these halos (see Figure \ref{formtimes}).
Finite numerical resolution limits concentrations to below a few times $10^5 \rho_{\rm crit}$
especially in the smaller satellites (the constant upper percentiles within 100 kpc are 
artificial). Dashed lines are fits (eq. \ref{concVSrhobg}) to the percentiles measured 
beyond 100 kpc. Likely dwarf galaxy host halos (triangles, same as in Figure \ref{tracks3vAccr}) 
also follow the general relation.
}
\label{concVSr}
\end{figure}

The large number of subhalos in Via Lactea allows us to study their
abundance, distribution, and concentrations as a function of distance
from the Galactic center. For this analysis we use the ten spherical
shells plotted in Figure \ref{LagRadii}, each at all times containing
a mass of $0.3 \times 10^{12} \msun$ and centered on the main galaxy 
progenitor. Figure \ref{massfrEv}
shows the number of subhalos and the substructure mass fraction of each 
shell as a function of time.  The amount of substructure
is very closely linked to the formation history of its host: in
each shell the number of subclumps peaks between the epochs of turnaround and
stabilization. Some time after a shell has stabilized, the abundance of subhalos becomes
nearly constant. Most of the tidal mass loss at a given radius
happens early, during a relatively short epoch when the corresponding
shell is near the end of its collapse and approaching
stabilization, and most subhalos are experiencing their first pericenter 
passage (see \S \ref{peri}). At low redshift,  
the mass in substructure lost from tidal effects or clumps orbiting out of the shell
is approximately replaced by other clumps streaming into the shell. The number of 
satellites with $\vmax>5 $km/s within the final $M_{200}$ (ie. within all six inner 
shells) remains within 25 \% of its value at $z=0.5$ at all times until the present. 
This nearly constant substructure abundance over the last 5 Gyrs
seems to be at odds with the recently found trends that old virialized systems 
are less clumpy \citep{Gao2004,vdB2005,Zentner2005}. However, we too see such a trend for the 
larger subhalos: the number of subhalos with $\vmax>10\,\kms$ within the final $M_{200}$ 
decreases steadily from 159 at $z=0.5$ to 112 at $z=0$. It seems that the larger
subhalos (roughly within three decades in mass of the host) are a transient 
population that declines continuously after the host has stabilized
\footnote{\citet{Zentner2005} do indeed point out that this trend is stronger for more massive subhalos.}.
On smaller scales, however, substructure appears to be more persistent and less
dependent of the age of the host (cf. \citet{Taffoni2003} and \S \ref{section:survial}).

Today, the number of subhalos with $\msub > 4\times 10^6\,\msun$ is smaller in shells 
closer to the Galactic center, in agreement with previous studies
\citep[e.g.][]{Ghigna2000,Diemand2004sub,Gao2004}.
The $z=0$ number density of mass-selected
subhalos is well described by a cored isothermal profiles with a scale radius similar to 
that of their host, as proposed by \citet{Diemand2004sub} and the ratio of subhalo 
number density and host matter density is simply proportional to radius:
\begin{equation}
\frac{n_{\rm M0}  (r) }{ \rho_{\rm host} (r) } \propto 
r~~~~{\rm for}~~0.1<r/\rvir<1.0\;.
\end{equation}
This ``spatial antibias" becomes smaller when subhalos are selected by their present $\vmax$,
In this case the bias scales with enclosed host mass:
\begin{equation}
\frac{n_{\vmax0}  (r) }{ \rho_{\rm host} (r) } \propto 
M(<r)~~~~{\rm for}~~0.1<r/\rvir<1.0\;  .
\end{equation}
The velocity dispersions of these spatially extended samples are larger than the dispersions
of the more concentrated, dark matter component, in good agreement with stationary
solutions of the Jeans equation applied to this two component system 
(as in \citealt{Diemand2004sub}). The orbital anisotropy parameter 
$\beta=1-0.5 \sigma_{\rm tan}^2/\sigma_{\rm rad}^2$ is identical for both the subhalos and the dark
matter: $\beta(r) \simeq 0.55 (r/\rvir)^{1/3}$ for $0.2<r/\rvir<1.0$.
Tidal mass loss, which causes the spatial and velocity distributions of these subhalo
samples to differ from the dark matter distribution, does not alter the 
orbital properties of substructure. Locally, we find a value of $\beta( r=8\kpc) = 0.12$.

While today the distribution of satellites is more extended than the dark matter,
the opposite is true at high redshifts: the inner shells are much more clumpy.
At $a=0.1$ even the smallest halos considered here
($4.0\times10^6 \msun$) correspond to rare density fluctuations (about
2$\sigma$). The enhanced subhalo abundance in inner shells at $a=0.1$
thus reflects the bias of high-$\sigma$ peaks towards the centers of
larger scale fluctuations \citep{Cole1989,Sheth1999}.

Figure \ref{slopeEv} shows the best-fit power-law slopes of the substructure
cumulative velocity function in the range $5-50\,\kms$ and the cumulative mass 
function in the range $4.0\times10^6-4.0\times10^9\,\msun$. The innermost shell (top left)
is affected by numerical resolution effects and small subhalo numbers.
The slopes show no strong trends with time (as in \citealt{Gao2004,Reed2005}) 
or distance from the main progenitor.
There is a slight trend toward steeper mass and velocity functions in the inner shells,
and toward a steepening with time from the epoch between turnaround and stabilization.
These small trends are suggestive of tidal mass losses being more significant 
in more massive systems and leading to steeper mass and velocity functions 
(see \S \ref{section:survial}).

The evolution of the median concentration is shown in
Figure~\ref{concEv}. The concentration parameter is related to the
cosmic mean density at the halo formation time
(\citealt{Navarro1997,Bullock2001conc,Wechsler2002}). Early forming halos,
which are well resolved in our simulation, do indeed
have high concentrations in all radial shells.\footnote{With our
physical (non-comoving) definition of concentration
(eq. \ref{eq:conc}), early forming halos have high
concentrations from the collapse redshift, whereas their 
$\cvir$ parameter would grow with time
to reach high values only recently.}. In the outer shell the median
concentration decreases with time because of the continuous formation of
new, lower concentration halos. In the inner shells the downward trend
is halted as the shell collapses and the abundance of subhalos freezes 
(cf. Figure~\ref{massfrEv}).  Note how the level of this
floor, between turnaround and stabilization of a shell, lies at higher
concentrations for the inner shells as they turnaround at higher redshifts.
After a shell stabilizes, the median
concentration grows because tidal forces remove mass from the outer
subhalo regions, thereby reducing $\rvmax$ and increasing the
mean density within this radius, $c_V$ \citep[][see also
Section~\ref{tracks}]{Kazantzidis2004}.
Together with the median, the 68\% scatter in subhalo
concentration is also growing.  This may be caused by the increasing
amount of mixing between newly infalling, low-concentration halos
and strongly stripped, high-concentration clumps.

At $z=0$ we find a clear trend for higher concentrations closer to the
halo center, as shown in Figure~\ref{concVSr}. We use the following
simple empirical fit to approximate this relation:
\begin{equation}\label{concVSrhobg}
c_V (r) = a \left[\frac{\rho_{bg}(r)}{\rho_{\rm crit,0}} \right] ^{b} \; ,
\end{equation}
where $\rho_{bg}$ is the average density of the corresponding spherical shell around the main host.
The best-fit coefficients are $(a,b) = (5895, 0.33)$ for the median concentrations,
(2997, 0.16) for the lower (16th) percentile, and (19370, 0.39) for the upper (84th) percentile.

The higher formation redshift of the inner halos is not enough to fully
explain this concentration-radius relation. If we simply assume that the
median halo concentrations are proportional to the mean cosmic density
at the median formation epoch of these halos (dotted line in
Figure~\ref{concVSr}) we do indeed get a qualitatively correct trend
with radius, but the effect is not strong enough. Tidal interactions
must significantly contribute to the final concentration versus radius
relation. In \S\ \ref{section:avgtracks} we confirm that this is indeed
the case: many halos did pass through the inner regions of the host at some 
earlier time and lost significant mass from tidal
stripping. Interestingly, tides seem to increase the median
concentrations (and scatter) even beyond $r_{200}=389$ kpc (shells
number 7 and 8). To summarize, the concentration-radius relation
is caused by the combined effect of two different processes:
\begin{itemize}
\item[i)] The formation of new small-scale structure stops when a
shell collapses. Inner shells turn around and collapse earlier, and
therefore contain earlier forming subhalos with a higher median
concentration.
\item[ii)] Tidal interactions within the host halo increase 
subhalo concentrations.
\end{itemize}

\section{Evolutionary tracks of subhalos}
\label{tracks}

The parallel group finder 6DFOF \citep{Diemand2006,Diemand2007} finds
peaks in phase-space density, i.e.\ it links the most bound particles
inside the cores of halos and subhalos together. The same objects
identified by 6DFOF at different times therefore always have quite a
large fraction of particles in common. In most cases this fraction is
over 90\% between two subsequent Via Lactea snapshots (separated
by 68.5 million years). This makes finding progenitors or descendants
rather easy. When tracing halos backwards in time, we link a halo ``A''
to its main progenitor ``B'' only if A contains at least 50\% of
the particles in B and if B contains at least 50\% of the
particles in A. This definition is time symmetric and we use the same
links when we follow halo histories forward in time. When a (sub)halo
merges with a larger group its forward history ends with a special
merger flag that points to the ongoing track of the merger
remnant. We include in our analysis only halos larger than $\vmax=5\,\kms$ 
at some point during their history. These halos
are resolved sufficiently well so that one can follow both their smaller, high-redshift progenitors
and also their present-day remnants, even if they did suffer large tidal mass loss.
The well resolved sample selected this way contains 3883 halos,
i.e.\ it is large enough to offer good statistics.
Starting at $z=0$ we identify the main progenitors of all such halos in each snapshot back
to at least $z=10$, when some progenitors start to become to small to be
resolved and identified with 6DFOF. The dotted lines in the right hand
panels of Figures \ref{avgtracks} and \ref{avgtracks_outer} show the
fraction of our halo sample for which we found a main progenitor as a
function of time.

\subsection{Density profiles during tidal mass loss}
\label{subsection:profileEv}

The evolution of the mass distribution in satellite halos undergoing
tidal stripping is often studied within an external fixed potential
\citep[e.g.][]{Dekel2003,Hayashi2003,Kazantzidis2004,Read2006}. The resolution tests in
\citet{Kazantzidis2004} show that numerical effects lead to
significant additional mass loss when an infalling subhalo is resolved
with N=0.5$\times 10^6$ particles and stripped down to a few thousand
particles within a strong tidal field.  The biggest subhalos in Via
Lactea are almost as well resolved as the high resolution case in
\citet{Kazantzidis2004}. Many of the smaller ones lie far below their
low resolution example and will suffer from artificial mass loss,
especially when the tidal forces are strong, i.e.\ in the inner
halo. Here we concentrate on the response to the tidal forces at
pericenter passage of two of the largest, best resolved subhalos.

\begin{figure}
\epsscale{1.2}
\plotone{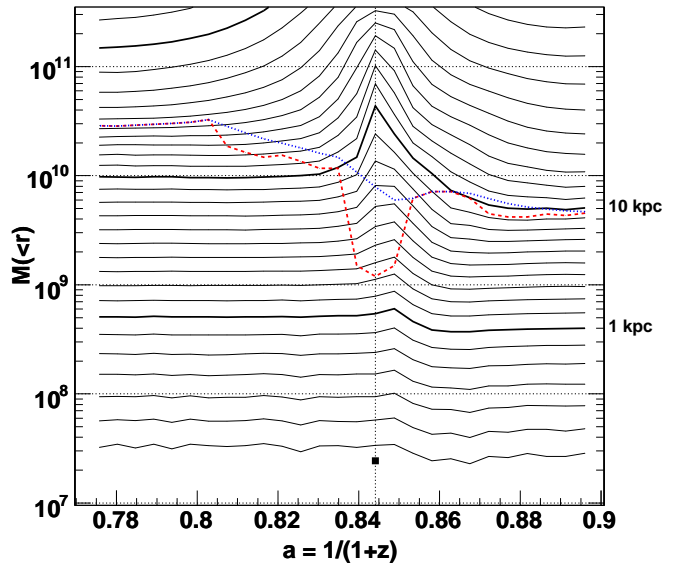}
\caption{Evolution of the mass distribution ({\it solid lines}) and
tidal mass ({\it dashed line}) of a subhalo undergoing strong tidal forces.
The thick solid lines show the mass enclosed within 1.0 and 10 kpc
spheres  around the subhalo center. The thin solid lines correspond
to the mass within nine intermediate radii (1.3,1.6, 2.0, 2.5, 3.2, 4.0, 5.0, 6.3 and 7.9). 
The smallest radius shown is 0.251 kpc, which is 2.8 times the force resolution.
This halo approaches  the Galactic center to within 8.3 kpc at $a=0.844$ 
({\it vertical dotted line}).
Tidal mass loss is larger in the outer parts, but also the inner subhalo loses mass. 
In response to the strong tidal shock the satellite contracts just after the pericenter
and expands soon after. 
The mass retained at $a=0.9$ is 3.9 times larger than the
mass within the tidal radius at the pericenter. A ``delayed" tidal mass 
({\it dotted line}) may be a better approximation to the bound mass. At $r_{\rm eq}=0.20$ kpc 
({\it square}) the subhalos internal orbital timescale matches the duration of the tidal shock. 
(see main text for details).}
\label{protracks2_17863}
\end{figure}

The first example is given is Figure \ref{protracks2_17863}. This
subhalo was accreted near $a=0.6$ and completes three pericenter
passages before $z=0$. Its full track is shown in Figure
\ref{tracks2vAccr}, while
Figure \ref{protracks2_17863} depicts the mass distribution around this
subhalo as it completes its second pericenter passage at
$a=0.844$ ($r_{\rm peri}$ = 7.0 kpc). 
The plot shows the mass enclosed within spherical windows of fixed
physical radii as a function of time. Not all of the enclosed mass
will be bound to the subhalo. At the peak of the 10 kpc line, for
example, the majority of enclosed mass is associated to the underlying host, 
the density of which is is $5.5\times10^4 \rho_{\rm crit} = 8.3 \times10^6
\msun/$kpc$^3$ at 7.0 kpc. On the other hand the host
contribution to the mass enclosed within 1 kpc is negligible compared
to the subhalo's own contribution. The brief increase in enclosed mass
for spheres with $r \simeq$ 1 kpc shortly after pericenter
passage reflects a temporary contraction of the subhalo as a response
to the rapidly varying potential, the so called tidal shock
\citep[e.g.][]{Gnedin1997}.  This contraction is only temporary, and
shortly afterwards the mass in the affected spheres actually
decreases. The energy input from the tidal shock results in a net
expansion \citep[e.g.][]{Hayashi2003,Faltenbacher2006}\footnote{The
expansion is large enough to overcome any tidal compression
suggested for subhalos orbiting in a $\Phi \propto r$
potential, as in the $\rho \propto r^{-1}$ inner region of the host
halo \citep{Dekel2003}.}.

Particles in the outer regions of the satellite complete only a tiny
fraction of their orbit around the subhalo center during the duration
of the tidal shock. For these particles the tidal shock is impulsive
and results in a maximal energy change. Particles near the subhalo
center are less affected, since their internal orbital period is
shorter than the duration of the shock. According to \citealt{Gnedin1997},
the energy input is proportional to
 \begin{equation}
 \Delta E(r)  \propto \left[ 1 + \omega(r) \tau \right]^{-5/2}\; ,
 \end{equation}
where $\tau=\pi r_{\rm peri} / V_{\rm peri}$ is the duration of the
tidal shock and $\omega=v_{\rm circ}(r)/2\pi r$ is the inverse of the
circular orbit time in the subhalo. 
For the small pericenter in this example the shock duration is only
$\tau=\pi$ (7.0 kpc)/(500 km/s) = 42.9 Myr.  This matches the subhalo
orbital time at $r_{\rm eq}=0.20$ kpc, i.e.\ at this scale $\omega(r)
\tau$ equals one and $\Delta E(r)$ is reduced to 0.18 of the maximal
value. Particles inside of $r_{\rm eq}=0.20$ kpc should be less
affected by the shock, unfortunately we cannot probe such small scales
reliably.  We do see however, that the shock is stronger in the outer
halo: the mass within 1 kpc drops to a new constant
value of 0.79 times the mass before the pericenter passage.  Farther
out the mass loss is larger: at $a=0.9$ the mass within 10 kpc is only
52\% of the value before the pericenter.  Therefore the remnants
of such a strong tidal interaction end up having lower densities at
most radii, but with steeper, more concentrated density profiles since
more mass is removed from the outer regions.

Before this pericenter passage the satellite has $\rvmax=7.5$ kpc. Its
tidal radius at pericenter is much smaller, only 1.6 kpc. According to
\citet{Hayashi2003}, satellites are fully disrupted when $r_t < 2 r_s
\simeq \rvmax$ at pericenter. This is clearly not the case in 
our example: the satellite survives this pericenter passage 
\footnote{It even survives the subsequent, closer pericenter at only 5.2 kpc.},
even though the tidal radius at pericenter is 4.7
times smaller than $\rvmax$. Figure \ref{tracks2vAccr} shows that in
general satellites survive even if they have several close pericenter
passages like the one studied here. In \S\ \ref{section:survial}
we find that total subhalo disruption happens very rarely, if at all.

\begin{figure}
\epsscale{1.2}
\plotone{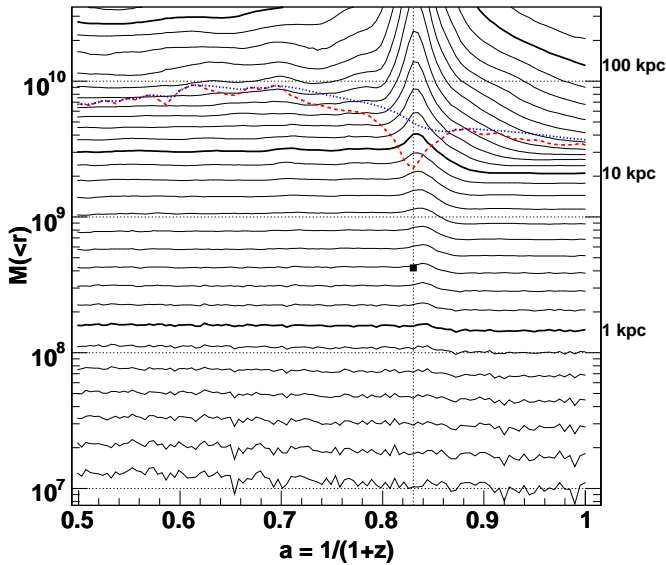}
\caption{Evolution of the mass distribution ({\it solid lines}) and
tidal mass ({\it dashed line}) of a satellite undergoing weak tidal forces.
This object falls into the
host and has only one pericenter passage at 56 kpc and $a=0.83$ ($z=0.20$).
The track of this subhalo is given in 
Figures \ref{tracks2vAccr} and \ref{tracks3vAccr} (dark green in the color version).
Tidal mass loss is large in the outer halo, while the inner subhalo
remains unaffected.}
\label{protracks2_31558}
\end{figure}

The second example is given is Figure \ref{protracks2_31558}. This
subhalo fell into the main host at $a=0.7$ and completes only one
relatively distant pericenter passage at $a=0.830$ ($r_{\rm peri}$ = 58.3
kpc).  Its track is shown in Figure \ref{tracks2vAccr}. The tidal forces at these
distances are much weaker, and as a result there is no significant mass
loss in the inner parts (less than 10\% within 1 kpc) and only
mild mass loss in the outer parts (29\% within 10 kpc). A tidal
shock can still be identified, but it is less strong and because of the
longer shock duration it does not reach as far in as in the previous
example. 
In this case $\tau=\pi (56 \kpc) / (423 \kms)= 406$ Myr and $r_{\rm
eq}=2.0$ kpc.  The increase at 10 kpc is purely due to the background
density that peaks at $2.1 \times10^5 \msun/$kpc$^3$ at
pericenter. The shells around 3 kpc do show some contraction and
expansion caused by the weak shock. Radii smaller than $r_{\rm eq}=2.0$ kpc 
are practically unaffected by the shock.
The mass retained at $z=0$ is larger (by a factor of 1.8) than the
mass within the tidal radius at pericenter: 10\% of this mass is
contributed by the host local background density at the 
position of the subhalo, and the remaining 90\% can be attributed to
the subhalo itself. Remarkably, we find that the fraction of particles
gravitationally bound to the subhalo (determined with
SKID\footnote{Available at
http://www-hpcc.astro.washington.edu/tools/ skid.html.}), is also 90\%.
Looking for additional bound material beyond the $z=0$ tidal radius
increases the bound mass of the subhalo slighly: by 28\% when going
out to 2.5 times the tidal radius. This larger bound mass is 16\%
higher than our tidal mass estimate, which includes the 10\%
contribution from the host halo.

The tests discussed above suggest that the simple tidal mass estimates
are a good approximation (within 20\%) to the bound mass at most
times. During pericenter passage our tidal mass may underestimate the bound mass
by factors of about 2 to 4. The same problem would affect subhalo
finders based on density only, since the bound material that is
missed by the tidal criterion by definition lies near (within a factor
of two) or below the host local background density.  As a result
subhalos at pericenter get assigned too little mass, and smaller
subhalos might be missed altogether. Our 6DFOF, in contrast, always finds
small groups, even when they lie below the background density. In the
current implementation, however, we also do not assign enough mass to
them.

The transient dip in tidal mass during pericenter passages seen in these two examples occurs because tidal stripping is not
instantaneous; many particles remain bound, even though they lie beyond the tidal limit when the subhalo
is near pericenter. Some semi-analytic subhalo models incorporate this effect by removing
only a certain fraction of the extra tidal subhalo mass at each time step $\delta t$:
\be
\Delta m_{d} = M(> r_t) \delta t / T_s \; .
\ee
$T_s$ is the time-scale for tidal stripping. We calculate the extra tidal mass
$M(> r_t)$ by subtracting mass that lies within the tidal radius in the current snapshot
from the ``delayed" tidal subhalo mass $m_{d}$ at the previous snapshot.
$m_d$ would be identical to the tidal mass in the limit of very rapid tidal stripping ($T_s \rightarrow \delta t$).
$T_s$ is often assumed to be equal to the satellite orbital time (e.g. \citealt{Taylor2001,Zentner2003})
\footnote{The time-scale used in  \citet{Zentner2005} is shorter: $T_{\rm orbit}/3.5$. A factor of $2\pi$
is missing in their Eq. 8 (A. Zentner, private communication)}.
Both of our (quite different) examples suggest that the stripping timescale is about
six times shorter than the time it takes the satellite to complete one full orbit ($T_{\rm orbit}$).
The delayed tidal masses assuming $T_s=T_{\rm orbit}/6$ are
shown with dotted lines in Figures \ref{protracks2_17863} and \ref{protracks2_31558}.
This means that mass loss can be relatively quick, e.g. it is possible for a subhalo to 
lose more than half of its mass during only a tenth of the time it takes to complete one orbit.

\subsection{The hosts of Milky Way dwarf satellites}
\label{subsection:dwarfhosts}

A promising scenario to explain the low numbers of Local Group dwarf
galaxies relative to the abundance of CDM subhalos
\citep{Moore1999,Klypin1999}, is to suppress star formation in small
halos below a filtering mass that increases after reionisation
\citep{Kravtsov2004}.  Similar models \citep{Bullock2000reion,Moore2006} select only
systems above the atomic cooling mass at the reionisation epoch 
($z\simeq$ 10). This too yields a realistic $z=0$ dwarf galaxy
population and the disrupted building blocks are shown to match the
spatial distribution and kinematics of halo stars around the Milky Way 
\citep{Diemand2005,Moore2006}.
In the Kravtsov et al.\ model the ten most luminous dwarfs are
practically all found in the subhalos that had the largest peak
circular velocities before these subhalos were affected by tides. The
mild time-dependence of the filtering mass only leads to a few
exceptions from this simple rule.

To illustrate the evolution of the hosts of dwarf galaxies in such
scenarios we select two samples consisting of ten objects. The
``largest before accretion'' (LBA) sample is made up of the ten
systems with the highest $V_{\rm max}$ throughout the entire
simulation. These ten systems all reached $V_{\rm max} > 37.3\,\kms$
at one time.  The ``early forming'' (EF) sample consists of the ten
halos with $V_{\rm max} > 16.2\,\kms$ (the atomic cooling limit)
at $z=9.6$. The EF sample corresponds to the \citet{Moore2006} model,
where sudden reionisation is assumed to have a strong effect on dwarf
halos. The LBA scenario corresponds to allowing star formation only
above a relatively high, constant critical size, a scenario of
permanently inefficient galaxy formation in all smaller systems, independently
of time-dependent changes in the environment like reionisation.  The
\citet{Kravtsov2004} model would yield a selection that is
intermediate between the LBA and EF samples.

\begin{figure*}
\epsscale{1.2}
\plotone{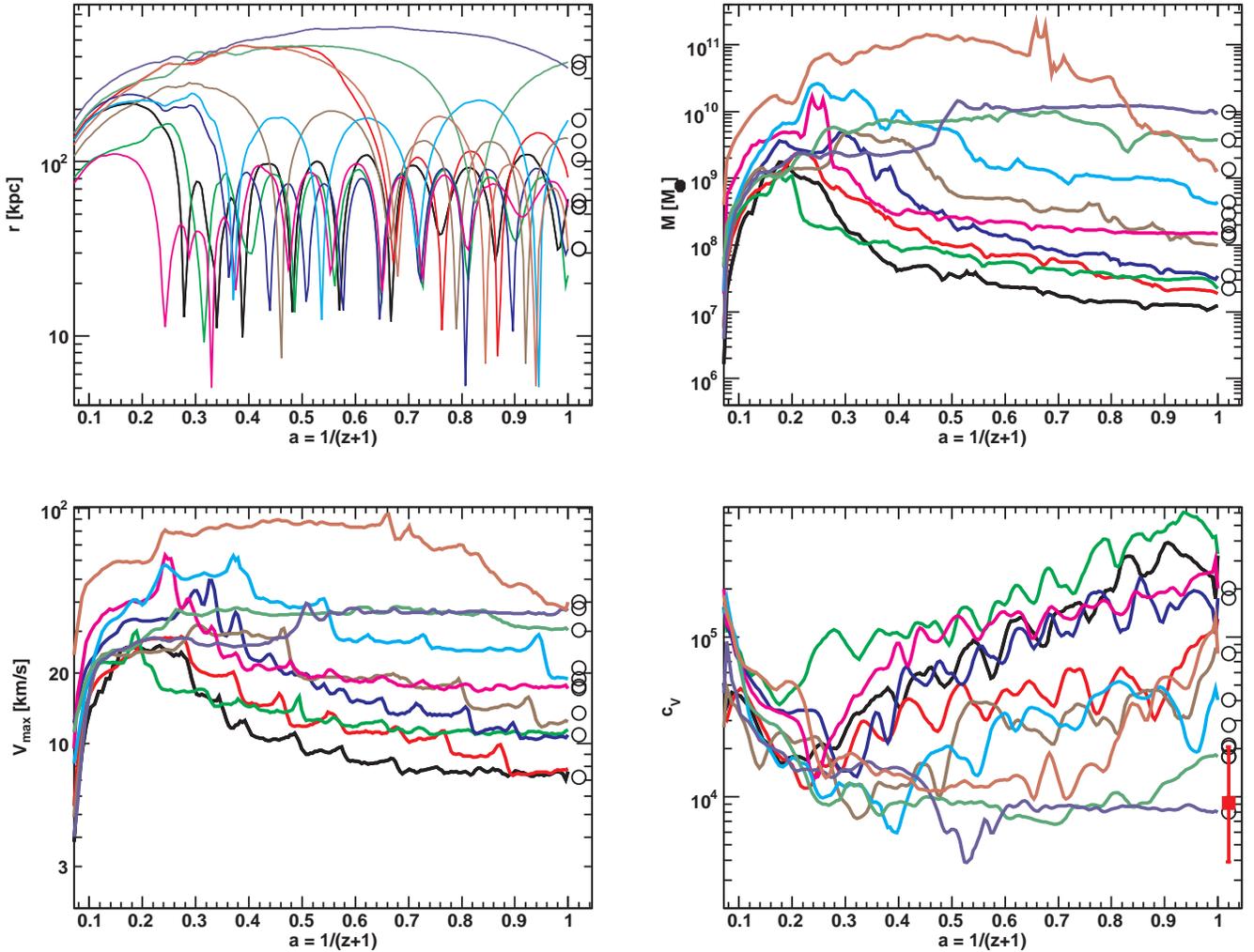}
\caption{Evolutionary tracks of the earliest forming substructure (EF
sample, {\it lines}) and $z=0$ properties of the subhalos with the largest
peak circular velocities before accretion (LBA sample,
{\it circles}).  {\it Top left panel}: distances to Galactic center versus 
cosmic expansion factor. Pericenter distances were calculated from 
the nearest snapshot (see \S \ref{peri}). {\it Top right}: evolution of tidal masses (or
$M_{200}$ for halos in low density environments, i.e.\ whenever
$M_{200}$ is smaller than the tidal mass). {\it Bottom left}: evolution of
subhalo peak circular velocities. {\it Bottom right}: evolution of
subhalo concentrations.  Square with error bars gives the median and
68 \% scatter of concentrations measured in field halos at $z=0$.}
\label{tracks2vAccr}
\end{figure*}

\begin{figure*}
\plotone{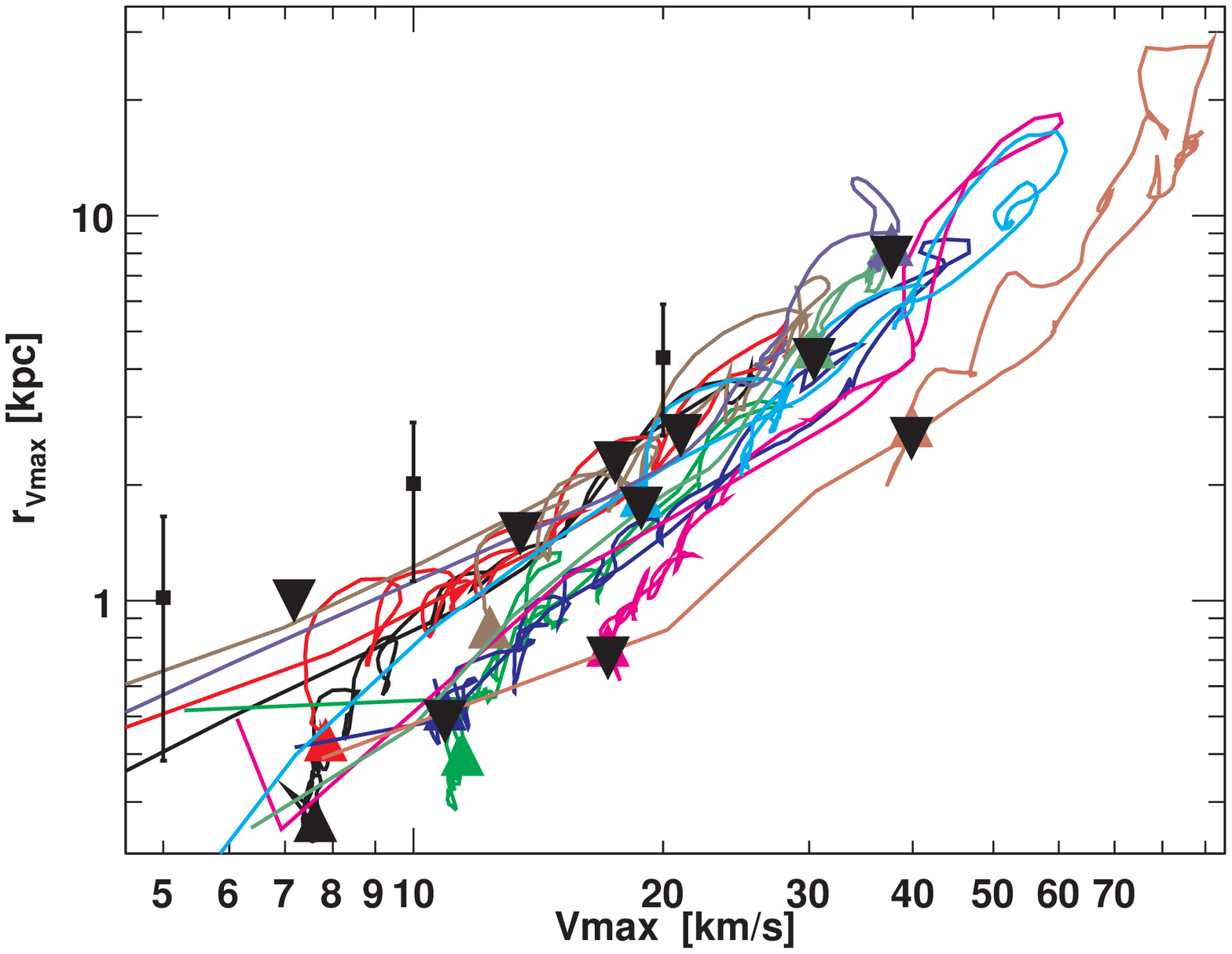}
\caption{Evolution in the $\vmax - r_{\rm Vmax}$ plane of likely dwarf galaxy host halos
(the EF sample, see text for details). Tracks start at high redshift
when halos are still small, i.e.\ in the lower left corner of this plot.
End points of the tracks at $z=0$ are marked with upward triangles. Only the end points
are shown for the LBA sample ({\it downward triangles}). Each halo tends
to grow and shrink within the same region of this plane, i.e.\ the end points
resemble their (relatively concentrated) high redshift progenitors.
Squares with error bars show the median location of $z=0$ field halos and
the 68\% scatter. Tidal stripping produces halos that are more
concentrated than those in the field (i.e.\ smaller $r_{\rm Vmax} $ at a given $\vmax$).
}
\label{tracks3vAccr}
\end{figure*}

Six out of ten objects turn out to be the same in both samples. Because
of the evolution of $\vmax(z)$ during the growth of halos at high redshift,
it is often the case that the ones reaching the largest sizes
before accretion are also the ones that have the largest
$\vmax(z\simeq 10)$. To avoid redundancy we plot the time evolution
only for the EF sample, but show the $z=0$ properties of the LBA
sample for comparison (Figures \ref{tracks2vAccr} and
\ref{tracks3vAccr}). The subhalo histories are obviously very
diverse. These halos have completed between zero and ten pericenter
passages. Some have lost over 99\% of their mass, others no mass
at all. The peak velocity $\vmax$ may have been reduced by up to a factor of six, or not at
all. Mass loss is often largest at the first pericenter passage (or the first
few) and then becomes smaller except for the largest subhalo, the orbit of which 
is decaying quickly because of dynamical friction \citep[e.g.][]{Taffoni2003}. All halos except one undergo pericenter
passages that can lead to tidally induced changes in galaxy
morphologies \citep{Mayer2001a,Mayer2001b}. For dwarf host 
halos found near the Galactic center today tidal interactions were more violent, happened more
often and started earlier, consistent with the very large mass-to-light ratios found for some
inner dwarf satellites \citep{Mayer2007}.

Concentrations decline with time while 
halos are growing. Later they remain constant for systems that lose
no or only little mass and end up in the range where field halo
concentrations are found (measured at $z=0$ between 1.5 and 4 times
$r_{200}$ using all systems with $\vmax$ between 6.6 and
$15\,\kms$).  For systems with large mass loss, on the other hand, the
concentrations increase with time and end up significantly above the
field halo range. Figure \ref{tracks3vAccr} illustrates the increase
in concentration during tidal stripping in the $\vmax-r_{\rm
Vmax}$ plane. With our definition of concentration $c_V$
(eq. \ref{eq:conc}), this parameter remains constant along $\vmax
\propto \rvmax$ tracks. In the $\vmax-\rvmax$ plane, halos start in
the lower left corner at high redshift and then wander quickly towards
the upper right corner. During this active mass-growth phase $\rvmax$
increases by a larger factor than $\vmax$, i.e. the concentration $c_V$ decreases.
After their active mass-growth phase they 
remain stationary in the upper right corner of this plane, until they experience tidal mass loss. 
Perhaps surprisingly, those
satellites that do undergo tidal stripping retrace their paths in the
$\vmax-\rvmax$ plane. This means that tidal stripping seems to
exactly undo the inside out subhalo assembly by removing mass from
the outside in. Each stripped down $z=0$ remnant ends up resembling
its own high redshift progenitor. Furthermore these $z=0$ remnants
have high concentrations, typical of high redshift systems, and
clearly higher than present-day halos that did not suffer significant mass
loss. Both samples follow the concentration-radius relation (see Figure \ref{concVSr}),
i.e.\ the highest concentrations are found in subhalos near the Galactic center.

The differences between the two samples are small and presumably hard
to detect from Local Group observations: at $z=0$ the four EF halos
that are not part of the LBA sample have smaller masses and lower
$\vmax$ but larger concentrations than the four LBA satellites that
aren't part of the EF sample. Unfortunately, current Local Group dwarf
galaxy mass models \citep[e.g.][]{Lokas2005,Kleyna2005,Koch2007} ,
based on radial velocity and surface brightness data, can only
place weak constraints on $\rvmax$
\citep{Strigari2006,Penarrubia2007}. In both samples the hosts of the
most luminous dwarf galaxies are significantly more concentrated than
field halos of similar mass: all ten $c_V$'s lie above the field
halo median, and most (nine for EF, seven for LBA) lie above the mean
halo scatter. Taking the higher concentrations of subhalos into
account is important when estimating the $\vmax$ values of dwarf
galaxy halos \citep{Strigari2006}; using field halo concentrations
instead \citep{Penarrubia2007} leads to higher $\vmax$ values.

\subsection{Ensemble-averaged evolutionary tracks}
\label{section:avgtracks}

\begin{figure*}
\plotone{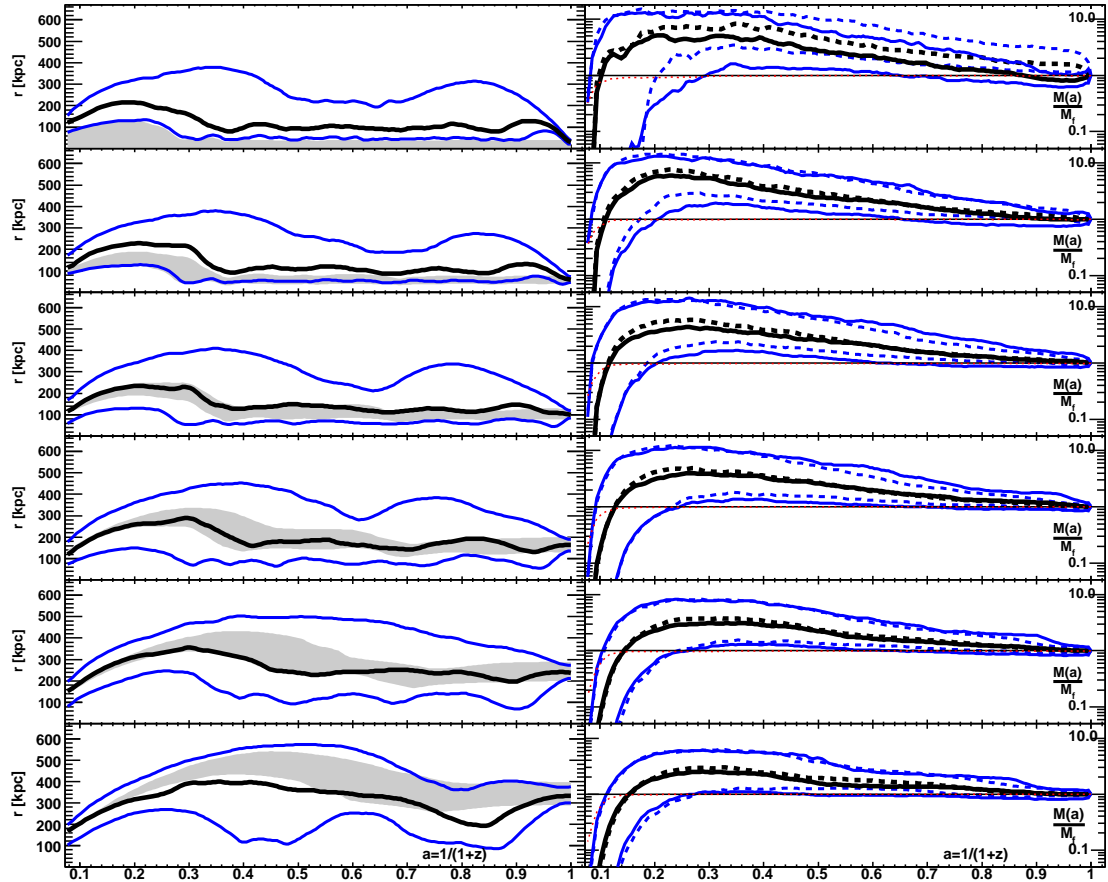}
\caption{Evolution of (sub)halos that end up in the same radial
shell at $z=0$.  Shells 1 to 6 are shown from top to bottom. Left
panels show median and 68\% scatter of (sub)halo distances from the
center of the host halo. Due to their radial orbits, subhalos are often found
outside of the fixed mass shell ({\it shaded region}) in which they end up at z=0.
Right panels show median and 68\% scatter of
the fraction of current mass to final mass ({\it dashed lines}) and
the one-fourth power of the fractions of peak circular velocity ({\it
solid lines}). The {\it thin dotted lines} show the fraction of halos
we are still able to trace at a given epoch. Halos without resolved
main progenitor are not included in the median radii, but they are counted
as zero when the medians of mass and peak circular velocity fractions
are calculated.}
\label{avgtracks}
\end{figure*}

\begin{figure*}
\plotone{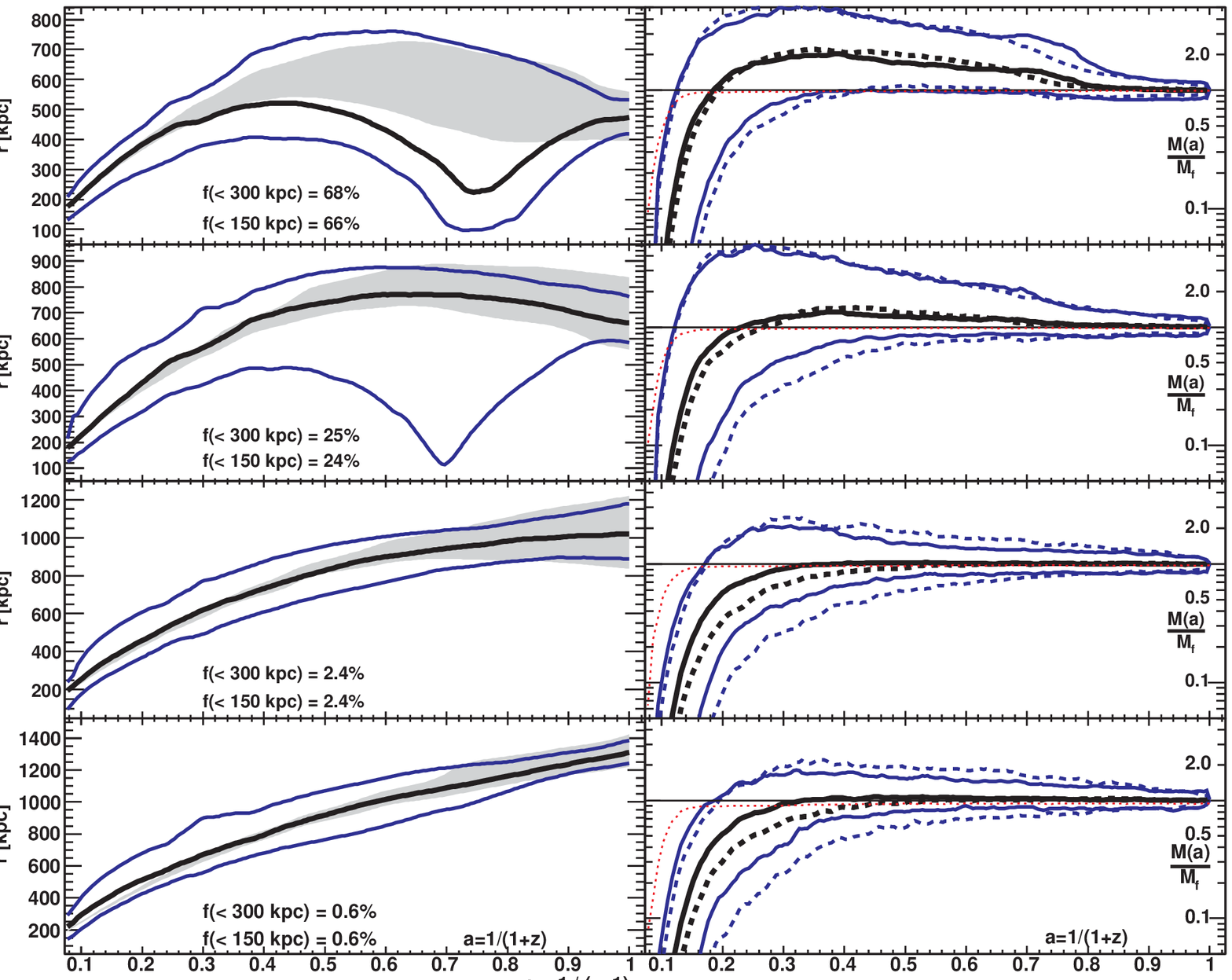}
\caption{Same as Figure \ref{avgtracks}, but for shells 7 to 10 that lie
beyond $r_{200}=398$ kpc at $z=0$. However, many of the ``field" halos
in shells 7 and 8 were actually ``subhalos'' at some earlier time. The left panels contain
include the fractions of halos
that approached the galaxy halo to within 300 and 150 kpc anytime after $z=1$.
Such former subhalos have a reduced mass (and $\vmax$), due to tidal interactions
with the main host at some earlier time. According to common
definitions of halo formation time these stripped former subhalos would
be classified as very early forming ``field" halos.
}
\label{avgtracks_outer}
\end{figure*}

After the few individual examples discussed above we now go on to
present ensemble-averaged evolutionary tracks, determined from a large
number of subhalos. The tracks are selected by two criteria:
the (sub)halo $\vmax$ must be at least $5\,\kms$ at some time, and the
object must end up in one of the ten spherical shells around the main
halo.

Figure \ref{avgtracks} show the median track of halos ending up within
the first six shells, i.e.\ within $r_{200}=389$ kpc. Median radii and
68\% scatter illustrate where most subhalos lying in a given
shell today were located in the past. Subhalos in the two innermost
shells at $z=0$ spend most of their time outside of these shells,
i.e.\ most of these satellites are on more extended orbits and are near
their pericenter at $z=0$.  The median radii in shells 3 to 6 are
roughly constant since $a=0.5$, i.e.\  the orbits of subhalos that lie
within one of these shells today were distributed roughly
symmetrically around this radial shell at earlier times. The scatter indicates
that the typical peri- and apocenters lie outside the final shell,
which is not surprising given the nearly isotropic orbits and 
the median ratio of apocentric to pericentric radii of 1:6 (see \S \ref{peri}).
 
The radii also reveal some regularities between current positions and
infall time. Halos in shell 1 today for example, are quite unlikely to
have fallen in around $a=0.6$ (and obviously also after
$a=0.9$). Those who did fall in around $a=0.6$ are most likely found
in shells 3 to 6 at $z=0$, after having completed one pericenter
passage (for an example, see the track near 400 kpc at $a=0.6$ in
Figure \ref{tracks2vAccr}).

In order to show how subhalo masses evolve, we
follow two mass indicators over time. One is simply
$M(a)/M(a=1)$, the ratio of the subhalo's mass at a given time to its
mass today. The other is based on the ratio of peak circular
velocities, $[\vmax(a)/\vmax(a=1)]^{4}$. The later definition is motivated
by the result that during tidal mass loss 
the peak circular velocity decreases roughly like $\vmax \propto M^{1/4}$ 
\citep{Hayashi2003,Kazantzidis2004,Kravtsov2004}. Whenever this approximate scaling
holds for our two mass indicators, they evolve proportional to each other.

We have plotted the median and 68\% scatter of these
two mass indicators versus scale factor in the right hand panel
of Figure~\ref{avgtracks}. In the outer regions near $r_{200}$ the two
do indeed agree nicely, but closer to the halo center the
masses are more strongly reduced than the peak circular velocities.
At least in the innermost bin the tidal masses seem to underestimate
the bound masses. From the median radii (left panel) it is clear that
most of the halos in this shell are at pericenter today, while in section
\ref{subsection:profileEv} we show that our definition of tidal mass
tends to underestimate the true bound subhalos mass at
pericenter. This explains the quick dip in the median mass fraction by
almost a factor of 2 near $z=0$.

Not surprisingly, both the median mass and $\vmax$ fractions show
clearly that tidal stripping is stronger near the halo
center. Stripping was also stronger at high redshifts: the median
$\vmax$ are roughly constant after $a=0.7$.  In shells 3 to 6 both the
median $\vmax$ and median radii are roughly constant after
$a=0.6$. This supports the findings of
Section~\ref{section:fixedmass}, that subhalo and host halo evolution
are closely linked. Early on the host system undergoes an active phase
of merging and mass accretion, during which the subhalos are accreted
and their mass is reduced quickly by tidal stripping at the first
pericenter passage (Section~\ref{peri}). Once the host halo has formed there is little
physical mass redistribution or accretion and the subhalo population
becomes stationary. During this quiet epoch most subhalos move on
stable orbits and their mass loss is relatively small.

\subsection{Formation histories and environment} 
\label{section:envdep}

Figure \ref{avgtracks_outer} shows the ensemble-averaged orbital
history of halos beyond $r_{200}$ today (in radial shells
7-10). According to their current location these would be considered
``field'' halos. However, many of them have actually orbited through
the host halo at some earlier time. The resulting tidal interactions
halted their growth and in many cases even reduced their mass and
$\vmax$. Such ``former subhalos'' would be classified as very early
forming field halos according to common definitions of formation time,
because they reached a given fraction of their current, tidally
reduced, mass or $\vmax$ much earlier than the average ``real'' field
halo that never passed through the host system. The population of
``former subhalos'' is significant: around galaxy clusters half
of all halos found between one and two virial radii today have
passed through the cluster at least once \citep{Balogh2000,Moore2004,Gill2005},
and for Via Lactea's subhalos this fraction is even larger, 
0.74. The fraction is very similar when only the most massive (sub)halo orbits
are used, i.e. we found no (sub)halo mass dependence.

\begin{figure}
\epsscale{1.2}
\plotone{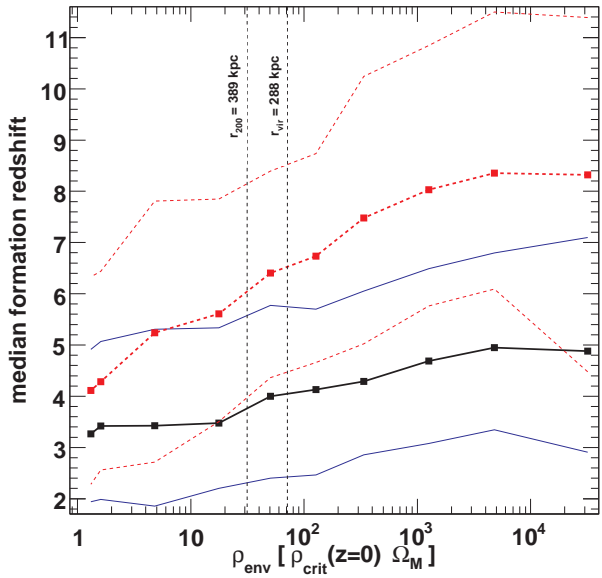}
\caption{Median $z_{\rm form}$ ({\it solid lines}) and $z_{85}$ ({\it dashed
lines}), both with 68\% scatter versus environment density.  Using
$z_{85}$, defined so that $\vmax(z_{85}) \equiv 0.85
\vmax(z=0)$, suggest that halos and subhalo living in denser
environments today (i.e.\ closer to the Galactic center) have formed
earlier. The relation holds at all environments probed here, even well
beyond the virial radius of the main halo.  Using the pre-stripping
halo size to define the formation time $z_{\rm form}$ the relation is
only significant within the virial radius.}
\label{formtimes}
\end{figure}

This illustrates clearly that a halo affects the formation histories of
many systems that are not within its virial radius (anymore). In
other words, the assembly history of CDM halos must depend on their
environment. Such correlations have indeed been quantified recently in
terms of stronger clustering of early forming sub-$M_*$ halos
\citep{Gao2005} and as earlier median formation times in high density
environments \citep{Sheth2004,Harker2006}. Both studies are based on formation
times defined relative to the $z=0$ halo properties, analogous to our
$z_{85}$ definition (eq. \ref{z85}). This definition of formation
time correlates strongly with mean environment density in and around
the Via Lactea halo, as seen in Figure~\ref{formtimes}, where we plot
median formation times in our ten radial shells (cf.
Figures~\ref{avgtracks} and \ref{avgtracks_outer}) as a function of
the mean density in these shells. 

On the other hand, for a formation time based on pre-stripping halo
properties, $z_{\rm form}$ (eq. \ref{zform}), this correlation
disappears for halos outside today's virial radius. It is maintained
inside the host halo because halos form before they are accreted and
median accretion redshifts are higher for subhalos that end up in the
inner shells (see Figure~\ref{avgtracks}). Note that the median
$z_{\rm form}$ lies below $z_{85}$ even in shells 9 and 10 where only
very few halo (2.4 \% and 0.006 \%, respectively) interacted with the
primary halo. This may be due to tidal interaction with a few other
relatively large ($\vmax \simeq 70\,\kms$) halos in the outskirts of Via
Lactea.

To summarize, we confirm the \citep{Balogh2000,Moore2004,Gill2005} result that
many subhalos end up in the field (according to common definitions)
and we illustrate that this leads to a clear environment dependence of
halo assembly histories like those found in
\citep{Gao2005,Harker2006}. A related explanation for the age
dependence of halo clustering based on tidal interactions with larger
halos was recently proposed by \citet{Wang2006}.  Defining halo
formation times relative to the maximal size a halo had over its
lifetime removes the environment dependence of median formation times.
Nevertheless, this does not change the \citet{Gao2005} conclusion,
that knowing only the $z=0$ mass of a halo is not sufficient to infer
its accretion history or halo occupation distribution in a
statistically correct way. Due to the strongly nonlinear tidal origin
of this effect it seems difficult to correct the analytical
approximations, suggesting that simulations should be employed
whenever structure formation needs to be followed accurately.

\subsection{Mass loss per pericenter passage}
\label{peri}

\begin{figure*}
\plotone{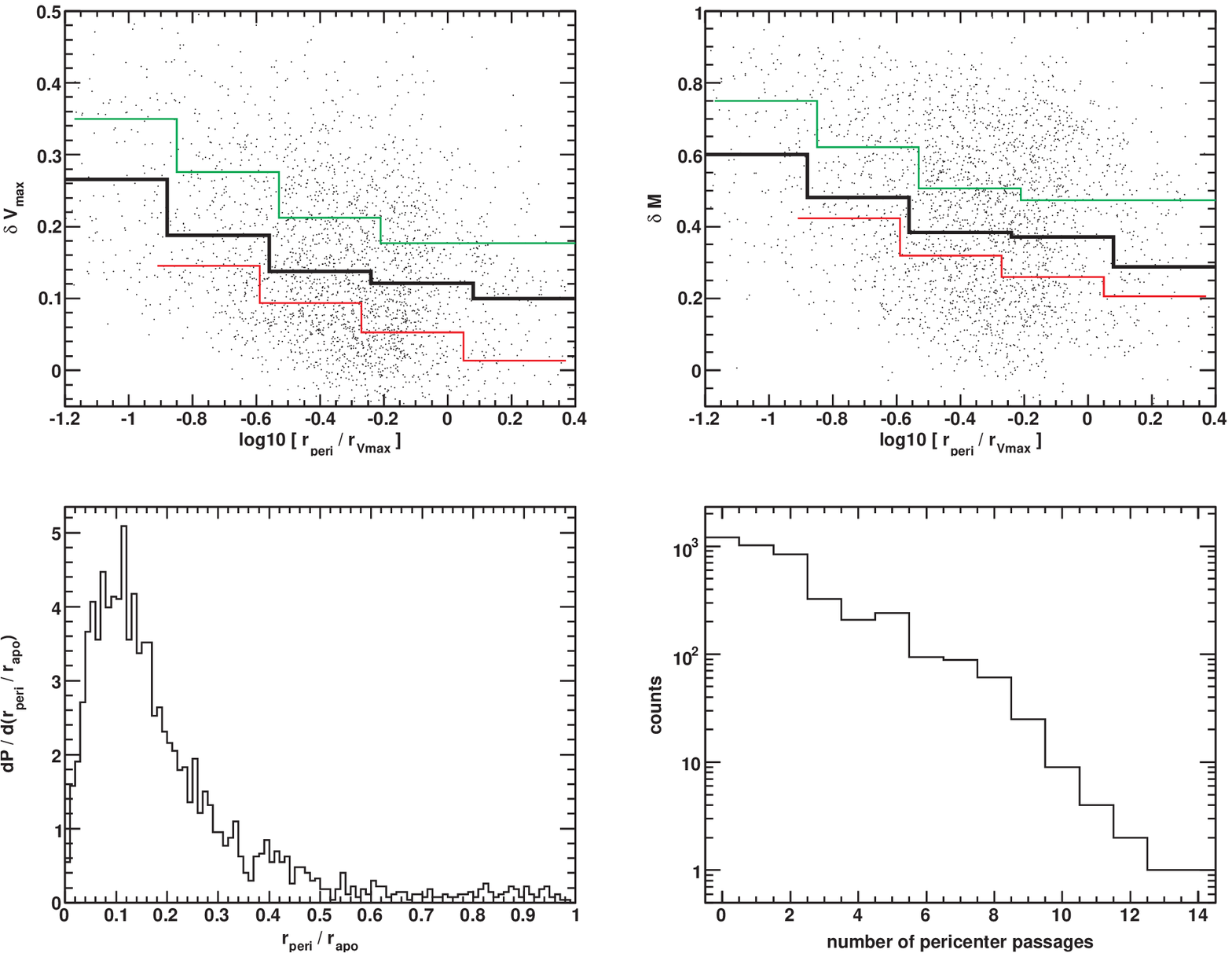}
\caption{{\it Upper left panel}: decrease in $\vmax$ per orbit versus pericenter distance.
{\it Upper right panel}: mass loss per orbit versus pericenter distance.
In both upper panels we show averages over all orbits ({\it thick solid lines}), over
first orbits ({\it upper thin lines}) and over the last orbit of those subhalos that
complete more than one orbit ({\it lower thin lines}).
Pericenter distances are normalized to $\rvmax(z_{\rm peri})$. Using a fixed scale
like $\rvmax(z=0)=69$ kpc gives very similar results since $\rvmax$ is roughly constant since the last
major merger at $z\simeq1.7$.
The lower left panel gives the ratios of pericenter distance to distance
at the subsequent apocenter and the lower right panel shows how many 
pericenter passages were completed by the 3883 subhalos in our sample.
}
\label{fig:peri}
\end{figure*}

In this Section we quantify how much mass a subhalo loses per pericenter
passage. We use the same sample of 3883 relatively well resolved ($\vmax>5$ km/s
at some redshift) subhalos as before. Pericenters are defined as local minima
in the distance to the main progenitor. Only minima within $4 \rvmax(z)$ are
counted, to exclude minima caused by orbits around other progenitors.
The time between stored snapshots (68.5 Myr) is too large to capture
all of the smaller pericenters. We calculate them by integrating
orbits using the position and velocity of the subhalo and the spherically
averaged mass distribution of the host halo at the nearest snapshot.
The 68\% interval of the directly measured $r_{\rm apo}/r_{\rm peri}$ distribution
is nearly identical (within 0.01) as the corrected one, i.e.\  for most
subhalos the calculated pericenter is similar to the one at the nearest snapshot.
Our subhalos made up to 14 such pericenter passages, but most of them
have completed only a few pericenters, or none at all (see Figure \ref{fig:peri}). 
The ratios of pericenter and
subsequent apocenter radii show that most subhalo orbits are quite radial. The median
ratio of our direct measurement (0.169) agrees very well with the derived value 1:6
from \citet{Ghigna1998}, who used $z=0$ subhalo positions and velocities and the
spherically averaged host density profile to integrate subhalo orbits approximately.
The 90\% interval extends from 0.035 to 0.666, i.e.\ only
5\% of all subhalo orbits are rounder than about 2:3.
These ratios are similar when we include only the last of several orbits (Table \ref{tab:peri}).
When we include only the first pericenter passages the ratios are slightly smaller, 
the median is 0.133. The anisotropy parameter $\beta(r)$ is
about $\beta(r) \simeq 0.55 (r/\rvir)^{1/3}$ for these subhalos (and also for the total the dark
matter) from $r/\rvir \simeq 0.2$ to 1.0. The positive $\beta$ values indicate
radially anisotropic subhalo (and dark matter) velocity distributions, and the anisotropy increases
with radius.

Masses (and $\vmax$) are measured at the apocenter after the pericenter passage ($t_f$),
and at an earlier time ($t_{i}$), so that the pericenter lies in the middle of these two times. 
This way we compare masses measured within similar, low background densities 
(even when a subhalo falls in for the first time) and we avoid the problem of underestimating
the subhalo mass at pericenter (see \S \ref{subsection:profileEv}). We do not use
late pericenter passages, i.e.\ when no subsequent apocenter is reached before $z=0$.
Mass loss is expressed as $\delta M \equiv [ M(t_i) - M(t_f)] / M(t_i)$ and the reduction in subhalo peak circular velocity
$\delta\vmax$ is defined accordingly. Figure \ref{fig:peri} shows that the average mass loss 
(and the decrease in $\vmax$) per pericenter is larger for orbits with smaller pericenter distances. The scatter is quite
large, the rms scatter is about 0.22 in each radial bin (0.11 for $\delta\vmax$).
The average mass loss
also depends strongly on the history of a subhalo: it is significantly larger when a subhalo passes 
through pericenter for the first time (Figure \ref{fig:peri} and Table \ref{tab:peri}).
The mass loss during the last of several orbits, on the other hand,
lies significantly below the average over all orbits. Many of the individual tracks in Figure \ref{tracks2vAccr}
illustrate this behavior, i.e.\ they show a large early mass loss and nearly constant masses (and $\vmax$)
near $z=0$. The effect also manifests itself in the larger average mass loss before $a=0.6$ in Figure \ref{avgtracks},
it is however smeared out because the first pericenters occur over a wide range of redshifts 
(the 68\% interval extends from $a=0.31$ to $a=0.63$), which is earlier but overlapping largely with the
distribution of all other pericenters (68\% within $a=0.48$ to $a=0.81$).
\begin{table}
\begin{center}
\begin{tabular}{cccc}
\hline
\hline
&all pericenters&1. pericenter& last of several \\
\hline
$\delta \vmax$ & 0.14$\pm$0.11& 0.22$\pm$0.11 & 0.10$\pm$0.08 \\
$\delta M$ & 0.41$\pm$0.22&0.58$\pm$0.20&0.31$\pm$0.18 \\
$\frac{r_{\rm peri}}{r_{\rm apo}}$ & $0.169^{\,0.395}_{\,0.070}$ & $0.133^{\,0.305}_{\,0.053}$ & $0.159^{\,0.336}_{\,0.078}$ \\
$\frac{r_{\rm  peri}}{\rvmax}$ & $0.25^{\,0.35}_{\,0.13}$ & $0.24^{\,0.35}_{\,0.12}$ & $0.27^{\,0.36}_{\,0.15}$ \\
$z_{\rm peri}$ & $0.67^{\,1.54}_{\,0.28}$ & $1.23^{\,2.20}_{\,0.61}$ & $0.27^{\,0.42}_{\,0.15}$ \\
\end{tabular}
\end{center}
\tablecomments{Mean and rms scatter are given for the decrease in $\vmax$ per orbit 
$\delta \vmax$ and the mass loss $\delta M$. For the other quantities the median
and the 68\% interval are listed.}
\vspace*{0.1in}
\label{tab:peri}
\end{table}

\subsection{Tracing survival and merging forward in time}
\label{section:survial}

The individual and ensemble-averaged tracks studied earlier by
definition only include halos that have survived (meaning they have a
remnant above our resolution limit) until today. This could
potentially bias the reported median mass loss rates to lower
levels. In this Section we quantify how many halos were stripped
below our resolution limit and check if this reduces the mass loss
reported for the survivors in the previous section.  We select halos
with peak circular velocities above $10\,\kms$ at $z=1$ and look for
their remnants today. Selecting only well resolved halos is necessary
for two reasons:
\begin{itemize}
\item[i)] tidal stripping and destruction are overestimated due to
numerical effects in barely resolved subhalos (the ``overmerging
problem'' \citealt{Moore1996,Kazantzidis2004}).
\item[ii)] one needs to identify the remnants even after severe tidal mass loss.
\end{itemize}

At $z=1$ there are 241 subhalos with $\vmax > 10\,\kms$ within a
sphere containing the final host halo mass (i.e.\ within shells
1-6). 232 of these are main progenitors of surviving subhalos, and two
merge into a larger surviving subhalo between $z=1$ and $z=0$. Only
the remaining seven subhalos are stripped below our resolution limit
and disappear from our $z=0$ sample. Half of the debris from the
tidally disrupted satellites are concentrated within a sphere of only
52 kpc around the Galactic center (for comparison, the half-mass
radius of the halo is 124 kpc), and the material beyond 52 kpc lies in
three tidal streams oriented towards the center. Both the
concentration of the debris and the radial direction of the streams
suggest that the destruction happened close to the Galactic center.
Extending the sample size by including all 1542 subhalos with $\vmax > 5\,\kms$
at $z=1$ yields similar results: 2.4 \% are lost and 1.3 \% merge into a larger
subhalo.

Since about 97\% of the subhalos selected at $z=1$ survive, the average
evolutionary tracks of surviving systems given in Section
\ref{section:avgtracks} are representative for the majority of
subhalos. It is also interesting to note that subhalo mergers are
extremely rare, between $z=1$ and $z=0$ the merger fraction is only about
1.3 \%.

\begin{figure}
\plotone{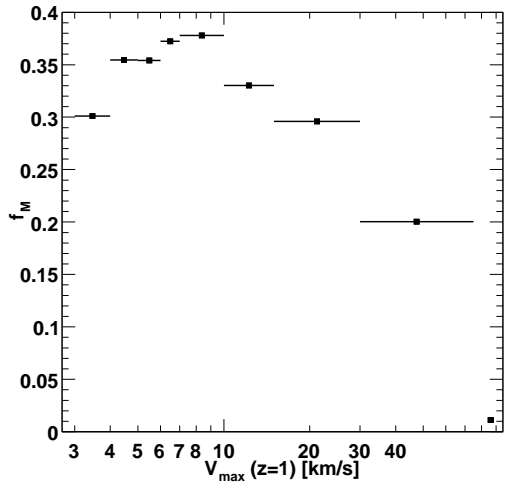}
\caption{Retained mass fraction from $z=1$ to $z=0$ versus peak circular
velocity range used to select the subhalos at $z=1$. The largest
subhalo is represented by the square at 86 $\kms$. Subhalos were
selected at $z=1$ within a sphere containing $M_{200} = 1.8 \times
10^{12} \msun$. The largest subhalos are sinking towards the center
due to dynamical friction, and they retain only a relatively small
fraction of their initial mass. Smaller subhalos are on more stable
orbits and lose less mass. The smallest systems (below about 7 $\kms$ or
1000 particles at z=1) suffer from an artificially large mass loss
caused by the finite numerical resolution.}
\vspace*{0.3in}
\label{surviving}
\end{figure}

Using the same subhalo selection we can also study what fraction of
their $z=1$ mass remains bound to their $z=0$ remnants. It turns out that
this fraction depends strongly on the initial mass range of the selected subhalos.
Larger subhalos retain less of their mass
(Figure~\ref{surviving}). The most massive halo (light brown track in
Figure~\ref{tracks2vAccr}) has an orbit that decays quickly due to
dynamical friction and it loses most of is mass (98.9\%) between
$z=1$ and $z=0$. 
Smaller subhalos are less affected by dynamical friction and lose
significantly less mass.  The increase in mass loss for subhalos with
$\vmax(z=1) < 7\,\kms$ is likely artificial, and caused by
insufficient numerical resolution. Note that the z=0 subhalo velocity
function of Via Lactea also starts to be affected by numerical limitations
below the corresponding z=0 scale of about 5 km/s (Paper I).

One consequence of larger mass loss in larger systems is that subhalo
mass and velocity functions should become steeper with time,
especially near when a region approaches virialisation, since this is
the time when most of the tidal mass loss happens
(Section~\ref{section:fixedmass}). Indeed, this trend can be observed
in Figure~\ref{slopeEv}. The mass range used in this Figure extends down to
$\vmax=5\,\kms$, the trend is stronger when only large subhalos are considered: the
slope of the cumulative velocity function
of subhalos with $\vmax > 10\,\kms$ measured within shells 1-6 grows
from 2.81 to 3.35 from $z \simeq 2$ to  $z \simeq 1$, and remains roughly constant from
then until $z=0$.

\section{Summary and conclusions}\label{section:concl} 

We have analyzed the co-evolution of the Via Lactea host halo and its
subhalo population. The simulation follows the formation of a Milky Way-size 
halo in a {\it WMAP} 3-year cosmology with 234 million DM particles. Here we 
summarize our main results:

\begin{itemize}

\item In agreement with \citet{Prada2006} we find that the formal
virial radius, defined in terms of comoving scales, underestimates
the actual virialized region of $\Lambda$CDM galaxy halos. The
increase in $M_{200}$ and $M_{\rm vir}$ after $z=1$ is almost entirely
due to apparent accretion, resulting from the artificial increase of the 
virial radius. Typically around 90\% of the final $M_{200}$ is
already within the final $r_{200}$ at $z=1$. When halo mass is based
on physical scales, such as $\vmax$ or mass within $\rvmax$, we find
no evidence for a late epoch of quiescent mass accretion as advocated
by recent studies \citep[e.g.][]{Wechsler2002,Zhao2003}.

\item The collapse factors of shells enclosing a fixed mass are very
different from the factor of two found in the idealized top-hat
collapse. This causes the shortcomings of $\rvir$. 

\item The abundance of substructure co-evolves closely with
the host halo. The subhalo mass loss rate
peaks between the epochs of turnaround and
stabilization and declines after a region has virialized. Mass
and velocity functions become slightly steeper during this process.

\item Tides remove subhalo mass from the outside in, which leads to
higher concentrations compared to field halos of the same mass. This
effect, combined with the earlier formation of inner subhalos, results
in strongly increasing subhalo concentrations towards the host center.

\item Selecting the earliest forming systems, or the largest before accretion, gives
largely overlapping and at $z=0$ nearly indistinguishable subhalo samples. They typically
show large, early mass loss and high concentrations, especially those found near the
Galactic center today.

\item We confirm the result by \citet{Balogh2000,Moore2004,Gill2005} that many
subhalos end up in the ``field'' (outside the virial radius) and
quantify the environment dependence of halo formation times caused by
this effect. Defining halo formation times relative to the maximum
circular velocity a halo reaches over its lifetime removes the
environment dependence of median formation times, but not the
environment dependence of halo mass assembly histories.  Due to the
strongly nonlinear tidal origin of the effect, correcting analytic
approximations seems difficult and simulations should be employed
whenever structure formation needs to be followed accurately.

\item At the first pericenter passage a larger average mass fraction is
lost than during each one of the following orbits. The median peri- to apocenter
ratio is close to 1:6 (as in \citealt{Ghigna1998}) and only 5 \% of the subhalo orbits
are rounder than 2:3.

\item We find that 97\% of all $z=1$ subhalos have a surviving $z=0$ remnant. The
retained mass fraction is larger for subhalos with smaller initial
mass. Satellites with $\vmax \simeq 10\,\kms$ retain about 40\% of their
$z=1$ mass at the present epoch.

\end{itemize}

\acknowledgments
It is a pleasure to thank Joachim Stadel for making PKDGRAV available
to us. We also like to thank the referee Andrew Zentner for
a very detailed and helpful report and Avishai Dekel, Vincent
Desjacques, Andreas Faltenbacher, Susan Kassin, Jason Kalirai, Andrey Kravtsov,
Ben Moore, Francisco Prada, Miguel Angel
Sanchez-Conde, Joachim Stadel, Simon White, and Marcel Zemp
for useful discussions and/or comments on earlier drafts of this paper.
J. D. acknowledges
financial support from the Swiss National Science Foundation and from
NASA through Hubble Fellowship grant HST-HF-01194.01 awarded by the
Space Telescope Science Institute, which is operated by the
Association of Universities for Research in Astronomy, Inc., for NASA,
under contract NAS 5-26555.
MK gratefully acknowledges support from the Institute for Advanced
Study.
P.M. acknowledges support from NASA grants
NAG5-11513 and NNG04GK85G, and from the Alexander von Humboldt
Foundation. All computations were performed on NASA's Project Columbia
supercomputer system.

{}

\end{document}